\documentclass{aa}
\usepackage[utf8]{inputenc}
\usepackage{natbib}
\usepackage{graphicx}
\usepackage{gensymb}
\usepackage{amsmath}
\usepackage{subcaption}
\usepackage{dblfloatfix}
\usepackage[colorlinks=true,allcolors=blue]{hyperref}
\usepackage[scientific-notation=true]{siunitx}
\usepackage[table,xcdraw]{xcolor}

\title{Iterative Angular Differential Imaging (IADI): An exploration of recovering disk structures in scattered light with an iterative ADI approach}

\author{L. M. Stapper\inst{\ref{inst1},\ref{inst2}} \& C. Ginski\inst{\ref{inst1},\ref{inst2}}}

\institute{Anton Pannekoek Institute for Astronomy, University of Amsterdam, PO Box 94249, 1090 GE, Amsterdam, The Netherlands \label{inst1} \and Leiden Observatory, Leiden University, PO Box 9513, 2300 RA, Leiden, The Netherlands \label{inst2}}

\date{\today}

\abstract
{ 
Distinguishing signal of young gas rich circumstellar disks from stellar signal in near infrared light is a difficult task. Multiple techniques have been developed over the years of which Angular Differential Imaging (ADI) and Polarimetric Differential Imaging (PDI) have been most successful. However, both techniques cope with drawbacks such as self-subtraction. To address these drawbacks we explore Iterative Angular Differential Imaging (IADI) techniques to increase signal throughput in total intensity observations.
}
{ 
This work aims to explore the effectiveness of IADI to recover the self-subtracted regions of disks by applying ADI techniques iteratively.
}
{ 
IADI works by feeding back all positive signal of the result from standard ADI over multiple iterations. To determine the effectiveness of IADI a model of a disk image is made and post-processed with IADI. Two versions of IADI are explored, classical IADI, which uses the median of the data set to reconstruct the point spread function (PSF), and PCA-IADI, which uses principal component analysis to model the PSF. In addition, masking based on polarimetric images and a signal threshold for feeding back signal are explored.
}
{ 
Asymmetries are a very important factor in recovering the disk due to less overlap of the disk in the data set. In some cases, a factor $\sim$75 more flux could be recovered with IADI compared to ADI. The Procrustes distance is used to quantify the impact of the algorithm on the scattering phase function. Depending on the level of noise and the ratio between the stellar signal and disk signal, the phase function can be recovered a factor 6.4 in Procrustes distance better than standard ADI The amplification and smearing of noise over the image due to many iterations did occur and by using binary masks and a dynamic threshold this feedback was mitigated, but it still is a problem in the final pipeline. Lastly observations of protoplanetary disks made with VLT/SPHERE were processed with IADI giving rise to very promising results.
}
{ 
While IADI has problems with low signal-to-noise observations due to noise amplification and star reconstruction, higher signal-to-noise observations show promising results with respect to standard ADI.
}

\begin{document}

\maketitle

\section{Introduction}



In recent years scattered light observations with facilities such as SPHERE \citep{Beuzit2019} and GPI \citep{Macintosh2014} have proven to be very successful in revealing the structures of circumstellar disks, in which planetary systems are created. Structures such as rings \citep[e.g.][]{Ginski2016, vanBoekel2017}, spiral arms \citep[e.g.][]{Avenhaus2017, Follette2017, Monnier2019}, asymmetric rings and gaps \citep[e.g.][]{Pohl2017, Benisty2018, Laws2020}, and shadows \citep[e.g.][]{Stolker2016, MuroArena2020} have all been observed with scattered light imaging (see \citealt{Benisty2022} for a recent overview of the field).

A key problem in optical or near infrared observations is that disk (or planet) signal is drowned out by the stellar speckle halo, which is several orders of magnitude brighter than disk scattered light or planet thermal emission. 
Different high-contrast imaging techniques can be used to remove the stellar light. Polarimetric Differential Imaging (PDI) uses the (un)polarized nature of the (star) disk, to disentangle star emitted light and disk scattered light \citep[e.g.][]{Kuhn2001, Apai2004}. However, not all light scattered by the disk is polarized. The polarization can vary up to 15\% depending on the scattering angle \citep{Min2005}, all the while depending on the dust grains and observation wavelength \citep{Murakawa2010}. Hence, to obtain the total scattered light intensity, a different technique is necessary.

Angular differential imaging, or ADI, is a high-contrast imaging technique used to remove stellar light from near infrared or optical images, while retaining the light from off-axis faint planets or circumstellar material \citep{Marois2006}. This is done by making use of field rotation; while the stellar point spread function (PSF) keeps the same position with respect to the telescope, off-axis faint planets or circumstellar material will rotate. ADI is in particular useful to retrieve the total intensity of the scattered light or the thermal emission of young planets (which is typically only marginally polarized, see e.g. \citealt{Stolker2017,vanHolstein2021}). Determining the total intensity is important to obtain information about the dust grain properties (e.g. \citealt{Tazaki2019}), planet emission (e.g. \citealt{vanHolstein2021}) and infer scatter angles in combination with polarimetric observations \citep{Ginski2021}. While ADI works very well on point-like sources such as exoplanets (e.g. HR~8799 \citealt{Marois2008, Marois2010}; 51~Eri \citealt{Macintosh2015}; HIP~65426 \citealt{Chauvin2017}), it struggles with extended structures because of `self-subtraction' due to parts of the structure overlapping during the observation. This leads to a number of problems:
 \begin{enumerate}
     \item Separation-dependent throughput: Due to the decrease in the length of the rotational arc with smaller separation from the central star, ADI throughput depends on angular separation. This leads to a suppression of disk signal in particular along the minor axis, which may turn ring-like radial sub-structures into "broken" arc-like structures. Importantly this will also significantly alter the scattering phase function and suppress signal for the smallest scattering angles.
     \item Spatial filtering: Self-subtraction depends on the spatial frequency of radial features. High spatial frequency features have a higher throughput than low spatial frequency features. This effectively results in a form of high-pass filtering, which is in particular problematic for young gas-rich disks, that tend to have a more diffuse morphology than debris disks which often are concentrated in sharper rings.
     \item Structure deformation: Self-subtraction can lead to non-trivial deformation of spatial features, in particular ring-like features may be deformed such that they can be misinterpreted for spiral structures.
     \item Detection bias: Finally self-subtraction in ADI strongly biases the global disk detection rate in scattered light toward exceptionally extended disks with high inclinations, while small or low-inclination disks may not be detected at all. 
 \end{enumerate}
 For a general discussion of these effects we refer to the overview given by \cite{Milli2012}. For specific examples, in particular of separation dependent throughput and spatial filtering, we refer to \cite{deBoer2016, Ginski2016, Perrot2016}. 


Recent works have explored different techniques in improving ADI to limit self-subtraction and thus increase signal throughput \citep[e.g.][]{Pairet2018, Pairet2021, Ren2018}. Our work explores a technique called Iterative Angular Differential Imaging (IADI). IADI aims to solve the issue of self-subtraction present in ADI to obtain similar results as obtained with PDI, but now using the full intensity of the scattered light. We are applying different data processing strategies to improve throughput. Specifically we combine the base concept of an iterative approach with classical ADI (cADI) and with principal component analysis (PCA) based ADI. The latter approach is identical to the GreeDS (Greedy Disk Subtraction) algorithm presented in \citet{Pairet2021}. We are expanding upon previous studies by including a data masking strategy based on PDI observations and by performing a detailed performance analysis of all techniques.

In \S\ref{sec:algorithm} the IADI reduction algorithm will be explained and in \S\ref{sec:model_setup} a simple scattered light model of a disk is made, which is in subsequent sections put through IADI post-processing. To quantify the recovery of the disk with IADI, \S\ref{sec:performance_metrics} two performance metrics are presented, with a particular emphasis on the recovery of the scattering phase function of the circumstellar disk (i.e. the scattering angle dependent brightness distribution of the disk). The effects of the elevation (i.e. distance above the horizon, related to the amount of field rotation) and inclination of the science target and the speed at which IADI converges to a particular final result is explored in \S\ref{sec:model_only}. The effects of the speckle field of the central star are analyzed in \S\ref{sec:model_and_star} and ways on how to suppress its affect on the reduction in \S\ref{sec:mask_and_threshold}. Finally, IADI is applied on three SPHERE observations towards circumstellar disks in \S\ref{sec:application_to_data}. Our results are summarized in \S\ref{sec:summary}.


\begin{figure*}[t]
    \centering
    \includegraphics[width=\textwidth]{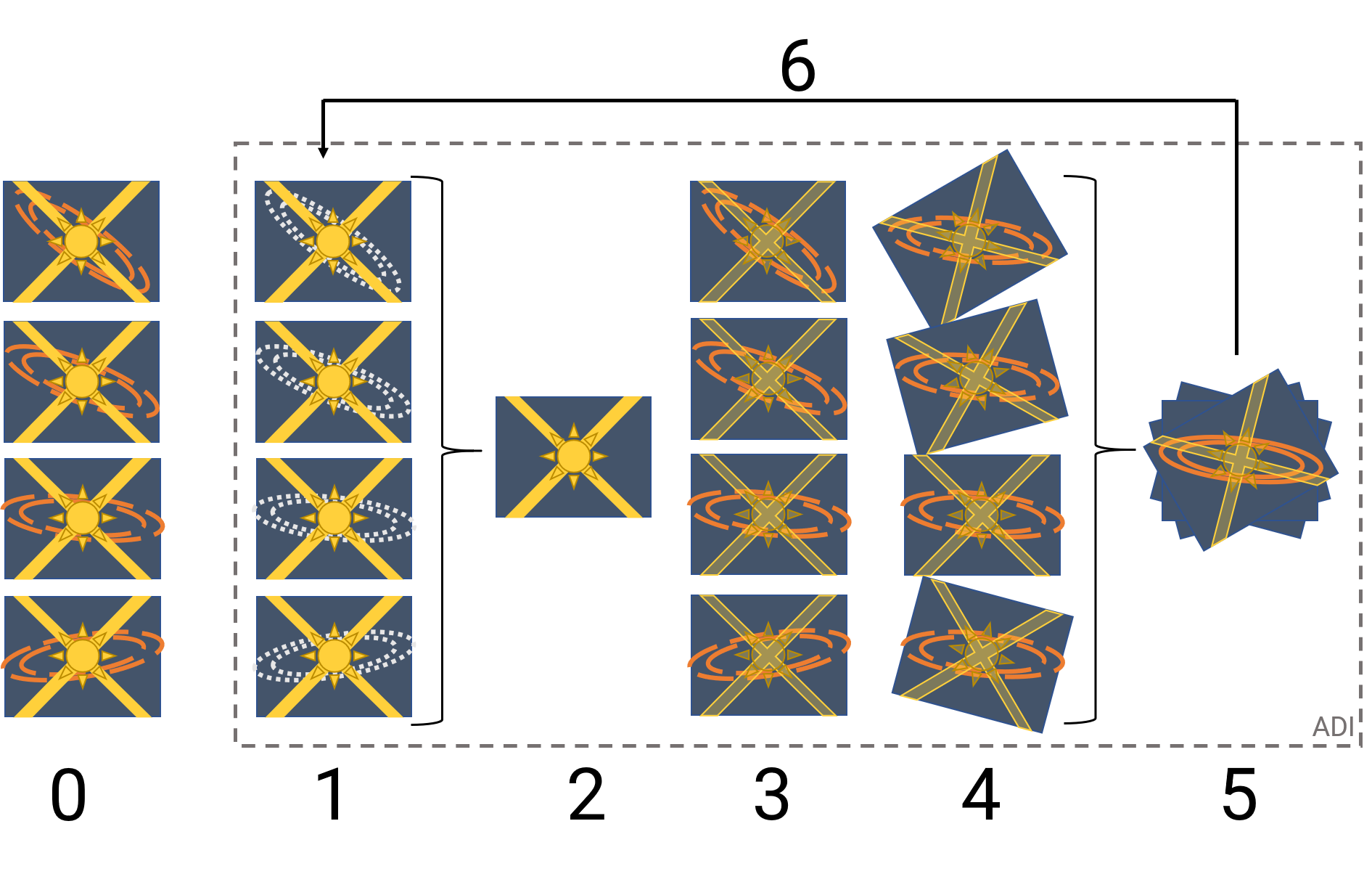}
    \caption{A flowchart depicting the different steps in the IADI reduction process. \textbf{0.} Copy of the original data set, which is used after every iteration. \textbf{1.} The original data set minus the final result from step 5. If this is the first iteration, the data sets of step 0 and 1 are identical. Even after some iterations, some disk signal is left in this data set after subtracting the final result. This is shown as the gray dotted ellipses. \textbf{2.} Take the median of the data set (cIADI) or construct the PSF with principal components after taking the mean of the data set (PCA-IADI). \textbf{3.} Subtract the found PSF from each image in the original data set copied in step 0. \textbf{4.} Derotate the data set such that the science object has the same orientation in all images. \textbf{5.} Take the median of all derotated images to combine them and get the final result. \textbf{6.} Feedback all positive signal of the final result from step 5 to step 1 and then repeat the process.}
    \label{fig:IADI}
\end{figure*}

\section{Algorithm}
\label{sec:algorithm}


ADI (and by extension, IADI) makes use of field rotation, which is the apparent rotation of an object in the field of view of an altitude azimuth mounted telescope due to the rotation of the Earth. Due to the point source like nature of the central star, the point spread function (PSF) of the star caused by the telescope (and observation conditions) will stay in the same orientation during the observation. On the other hand, any possible companion of the star such as a planet or a protoplanetary disk will rotate in the field of view of the telescope, see column 0/1 in Figure \ref{fig:IADI}. Hence, field rotation has the consequence of no change in orientation of the signal which needs to be removed from the image (the star) and a change in the orientation of the signal which contains the science object (the disk). Iterative Angular Differential Imaging expands upon ADI by doing it iteratively on the same data set, so part of the reduction process is the same as ADI (see the dashed box in Figure \ref{fig:IADI}). 

First (column 0 in Figure \ref{fig:IADI}), a copy of the original data set is made. In the first iteration, column 1 contains the exact same data set as column 0. The stellar PSF is isolated by taking the median of the data set along the time axis (i.e. along the pixels at the same position in each image in the data set). Ideally only the stellar PSF will remain, see column 2 in Figure \ref{fig:IADI}. Besides taking the median, principal component analysis (PCA, \citealt{Amara2012}) is also explored in this work. With PCA an image is decomposed into a set of functions, or principal components of the image, which, when combined into a linear combination, can represent the original image. This approach is implemented based on the \texttt{Python} package \texttt{\texttt{PynPoint}} \citep{Stolker2019}. In practice, first the data is mean subtracted and then reconstructed via PCA with a specific number of principal components. The resulting PSF found via either the median approach (classical IADI, or cIADI) or via PCA (PCA-IADI) is subtracted from the original data set, see column 3 in Figure \ref{fig:IADI}.\footnote{We note that we in both cases use the full science data set for the median combination as well as the PCA, i.e. we did not implement an "exclusion angle" as is done in some cases to limit self-subtraction (e.g. LOCI \citep{Lafreniere2007}, pyKLIP \citep{Wang2015}).} 

\begin{figure*}
    \centering
    \includegraphics[width=\textwidth]{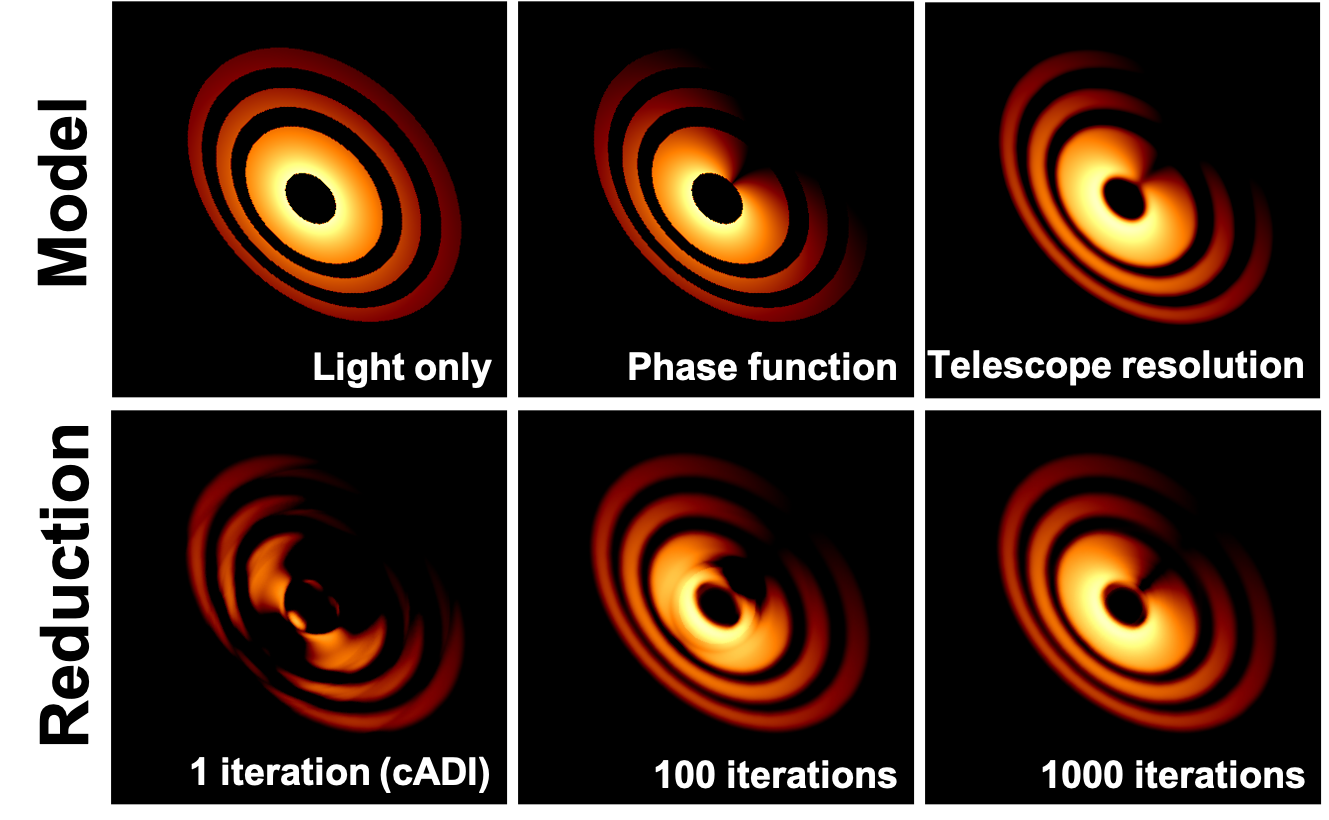}
    \caption{\textbf{Top panels}: The model used in this work. Shown with only a $\sim 1/r^2$ intensity dependence from the center of the disk, the addition of the phase function and telescope resolution. \textbf{Bottom panels}: The recovered disk after 1, 100 and 1000 iterations. This shows that with classical ADI large parts of the disk is not recovered due to self-subtraction, but that this can be recovered iteratively. All images use a log-scale to make the faint parts of the disk better visible.}
    \label{fig:model_layers}
\end{figure*}

After PSF subtraction, the images are rotated back, such that the science object in each image has the same orientation, see column 4 in Figure \ref{fig:IADI}. Then the images are combined by taking the median, see column 5 in Figure \ref{fig:IADI}. Now ideally all signal coming from the science object is recovered. However this is rarely the case, hence the iterative nature of IADI. As the bottom row of Figure \ref{fig:model_layers} shows, with classical ADI large parts of the disk is missing due to self-subtraction. Hence, in the final step all the positive signal is fed back by subtracting this from the original data set in column 0 and the process is repeated. In this way, column 1 contains less disk signal, so less signal of the disk will be in the recovered PSF and consequently less of the disk is subtracted in the final step. After many iterations, more of the disk is recovered (see the bottom row of Figure \ref{fig:model_layers}). Note that one can also feedback parts of the result instead of all positive signal with the use of a threshold or a mask, this is explored in Section \ref{sec:mask_and_threshold}. Asymmetry in the disk plays an important role in IADI (as it already does in standard ADI). The more asymmetric the disk is, the faster the process will converge to a final image.

\section{Model reduction}
\subsection{Model setup}
\label{sec:model_setup}

To test IADI, a model of a disk in scattered light is made. This model is 400$\times$400 pixels, each pixel having a physical size of 1$\times$1~au. Multiple rings are implemented, inspired by observations of disks (e.g. HD 97048 \citealt{Ginski2016}, RX J1615 \citealt{deBoer2016} and TW Hydrae \citealt{vanBoekel2017}), see panel a) in Figure \ref{fig:model_layers}. A specific inclination and rotation is achieved by using the \texttt{warpAffine} function from \texttt{OpenCV} \citep{opencv_library}. Young planet forming disks are still gas-rich and dust particles are stratified due to gas pressure along the vertical axis. Thus flaring is implemented in the model via an offset of the rings with respect to the center of the ring (see the detailed discussion in \citealt{deBoer2016}). For our disk model we are using the power law profile for the scattering surface height $H$ and the separation $r$ found by \citep{Ginski2016} for the disk around HD~97048. This profile describes the found flaring reasonably well up to a separation of $\sim 270$ au. Considering that the model disk used will have a separation of 200~au, this formula will be sufficient to simulate the disk height\footnote{We note that flatter disk profiles were recovered by \citet{Avenhaus2018} for several T~Tauri stars, which would lead to a slightly smaller offset of the disk rings from the stellar position.}. Illumination effects of the central star are implemented via a $\sim 1/r^2$ intensity dependence from the center of the disk, making the inner part of the disk brighter compared to the outer part. Moreover, the intensity also depends on the light scattering angle via the phase function. Because a physical model is outside the scope of this work, a `pseudo' phase function is implemented to mimic the same asymmetries in light distribution seen in observed disks via $I = \cos{\phi}$, where the intensity $I$ depends on the cosine of the azimuthal angle $\phi$. This pseudo phase function depends on the azimuthal angle instead of a scattering angle on which a real phase function would depend. This is making the part of the disk facing towards the observer brighter than the part facing away in a fairly simple way. Lastly, the model is put through a Gaussian convolution kernel from the \texttt{scipy} \texttt{ndimage} package \citep{2020SciPy-NMeth}, to remove sharp edges and give a finite resolution to the model. The final three steps are shown in the top row of Figure \ref{fig:model_layers}.

\subsection{Performance metrics}
\label{sec:performance_metrics}

To quantify how well the disk is recovered after IADI post-processing, two main metrics are used. First, in \S\ref{sec:model_only} only the disk will be processed by the IADI pipeline. Although this speckle-free reduction is not representative of the reality of observation conditions with real stellar noise, this simplified case allows to assess the impact of self-subtraction only on the disk. Second, in \S\ref{sec:model_and_star} the model is inserted into observations of a star, and thus cannot be directly compared to the model. Hence, the phase functions of the recovered disks will be compared to the phase function of the model via Procrustes analysis.

\subsubsection*{Retrieved flux comparison}
\begin{figure}[!t]
    \centering
    \includegraphics[width=0.5\textwidth]{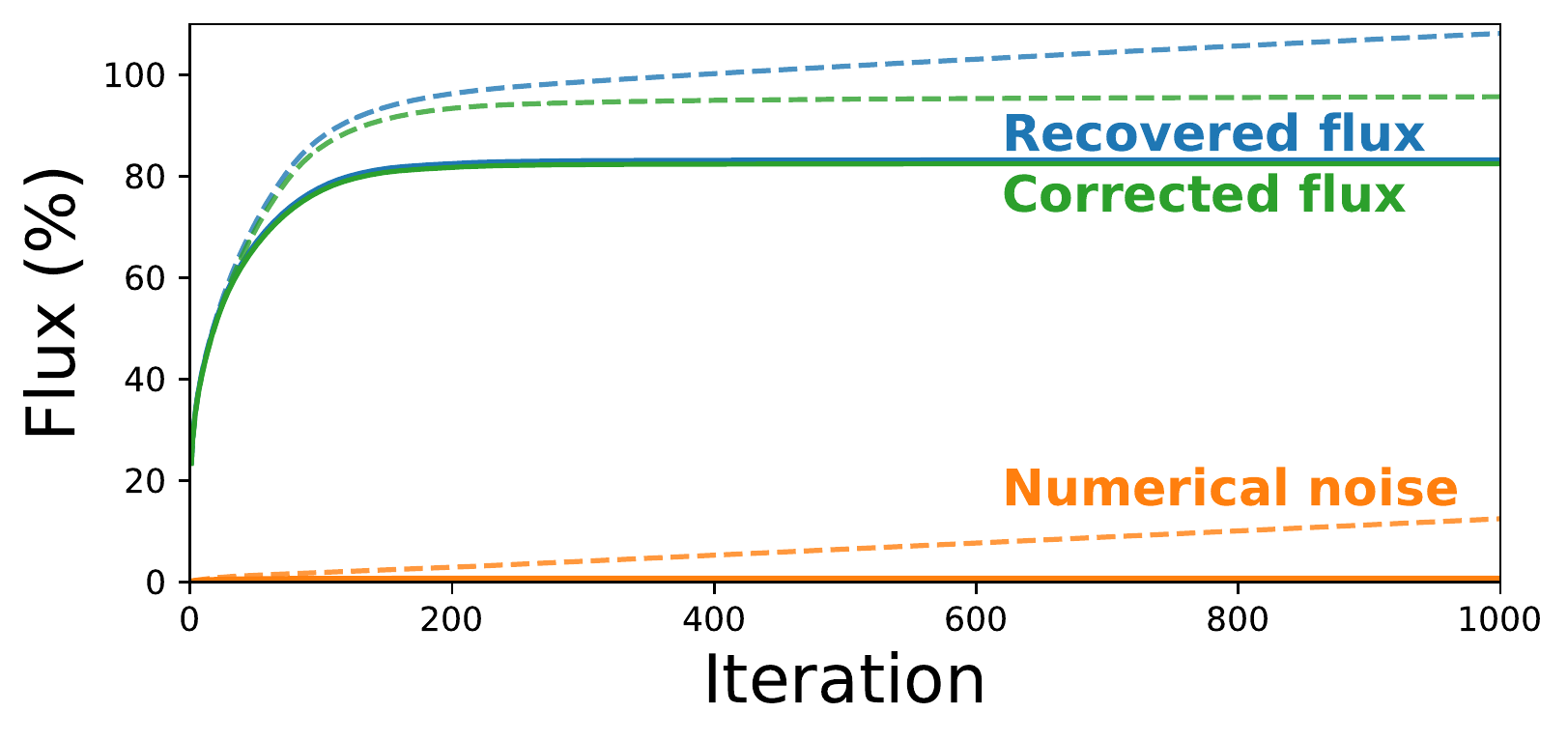}
    \caption{The recovered flux, numerical noise and corrected flux plotted against the number of iterations of a disk with an inclination of 45$\degree$ and elevation of 85$\degree$ processed with cIADI. Both spline interpolation (dashed lines) and nearest neighbor interpolation (continuous lines) are shown.}
    \label{fig:recov_num_corr}
\end{figure}

The first metric is the amount of retrieved flux compared to the inserted disk model. The original total flux of the disk is the sum of the values of each pixel in the original model. After every iteration this recovered flux is compared to the original value of the model. The recovered flux (in \%) is the ratio between the sum of the values of each pixel in the disk after some amount of iterations, divided by the sum of the values of each pixel in the original disk. In Figure \ref{fig:recov_num_corr} one can see that the recovered flux increases rapidly in the first $\sim200$ iterations, after which it increases linearly.

Contradictory to what one might expect, the total recovered flux increases to above $100\%$. This is due to numerical artifacts generated during the reduction process. The numerical noise is the noise introduced into the disk because of interpolation artifacts due to the many rotations during IADI processing. While in real observations this noise cannot be estimated, in this case the actual disk is known and thus can be used to estimate the amount of noise introduced. The percentage of numerical noise depends on the interpolation kernel used, see Figure \ref{fig:recov_num_corr}. After careful consideration, nearest neighbor interpolation was found to introduce the least amount of noise into the image while still recovering the majority of the disk, granted that enough images of the disk are present in the data set to still sample the disk well. Whether nearest neighbor interpolation is the most optimal choice should be considered on a case-by-case basis. The numerical noise is calculated by subtracting the original disk from the recovered disk and summing all positive values. In this way, every pixel where the value is larger than the original disk remains, which should come from numerical artifacts. However, everywhere where the disk had self-subtraction, the numerical noise introduced is not measured, so this only gives a lower limit on the numerical noise. Unfortunately, numerical artifacts will never be completely zero, but it can be made as low as possible by choosing the correct interpolation kernel (i.e. nearest neighbor interpolation for any further reduction done in this work).

Lastly, the corrected flux is the recovered flux corrected for the numerical noise. After correcting the flux for the numerical noise, the total amount of flux indeed does not surpass $100\%$, see Figure \ref{fig:recov_num_corr}.

\subsubsection*{Procrustes analysis}
Ultimately, the recovery of the phase function is an important goal after post-processing the data. Hence, a `pseudo' phase function is recovered by summing all positive pixels in bins in the azimuthal direction, which is then smoothed by a Butterworth low pass filter. To analyze the recovered shapes of these phase functions, Procrustes analysis is used \citep{Dryden2016}. The algorithm for comparing the recovered phase function and the model phase function works as follows. First, we translate the mean of the phase functions to the origin. Second, a uniform transformation is done via scaling the phase functions such that the root mean square of the distance to the origin is equal to one. Lastly, to quantify the difference between the two phase functions, the Procrustes distance is computed via

\begin{equation}
    d_{\text{Pr}} = \left(\sum_{i=1}^n \left[(x_i - u_i)^2 + (y_i - v_i)^2\right]\right)^{1/2}
\end{equation}

\noindent
in which ($x_i$, $y_i$) and ($u_i$, $v_i$) are the $i$-th coordinates of the recovered and model phase functions, respectively. The closer the Procrustes distance is to zero, the more similar the two phase functions are to each other. A Procrustes distance of $0$ means that, after removing translational and scaling factors, both functions have the exact same shape.

It should be noted that in \citet{Dryden2016} an extra step is mentioned in which any rotational differences are also removed. Considering that the phase functions are positioned in the same direction, this step was omitted. Furthermore, the models to which the recovered phase functions are compared to will be kept the same between all reductions. In this way all computed Procrustes distances can be compared to each other.

\begin{figure*}
    \centering
    \includegraphics[width=\textwidth]{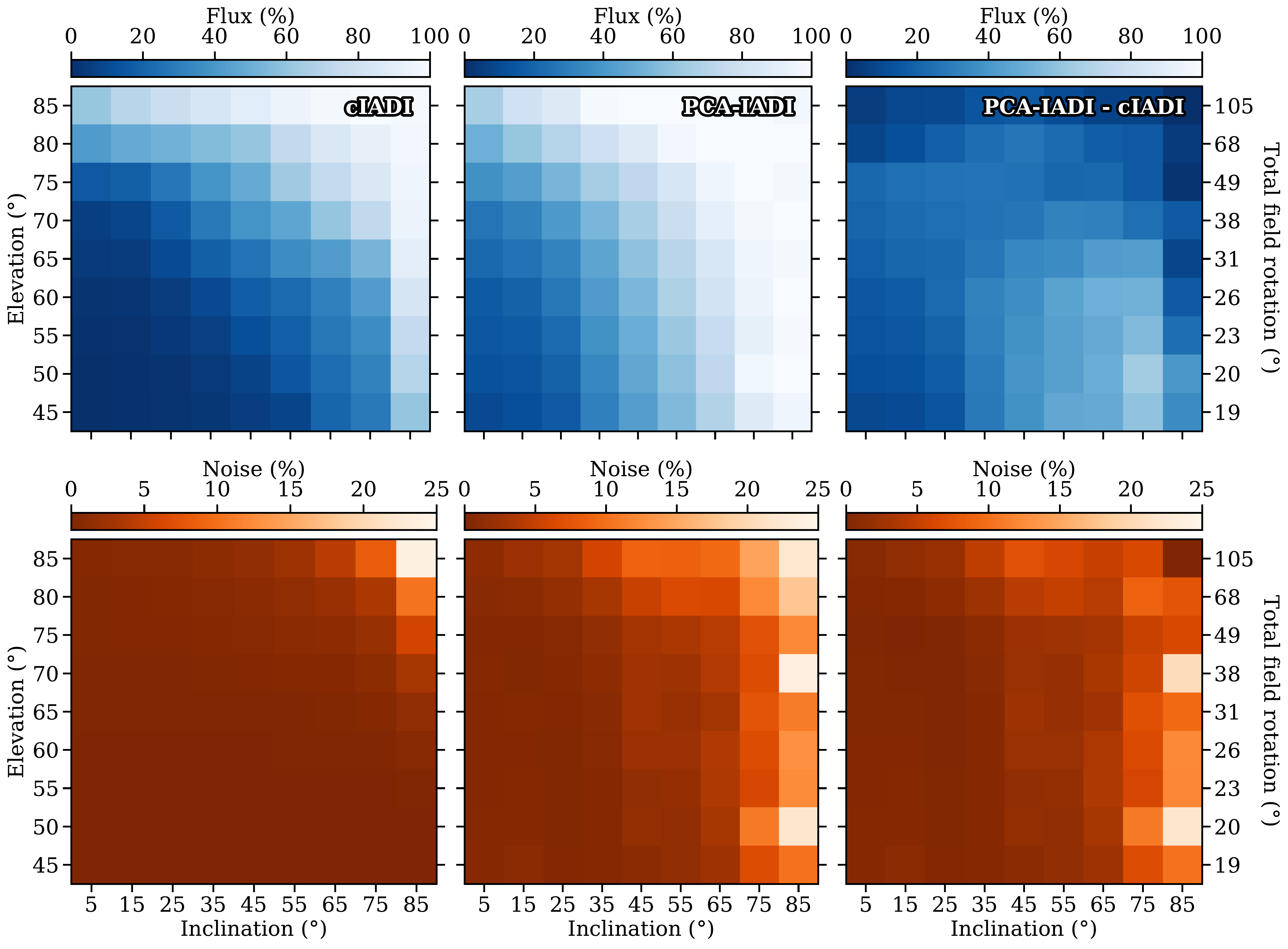}
    \caption{The top row shows the flux recovered by cIADI and PCA-IADI after 500 iterations for different inclinations and elevations/amount of field rotation. The bottom row shows the numerical noise generated during the iteration process. The right most column shows the difference between PCA-IADI and cIADI.}
    \label{fig:flux_matrices}
\end{figure*}

\subsection{Tests on disk model only}
\label{sec:model_only}




To asses how well the self-subtracted areas of the disk can be recovered with IADI, a 9x9 grid of models has been reduced for 500 iterations with cIADI and PCA-IADI for nine different inclinations and nine different elevations. The models are simulated at the Paranal observatory with the target at different declinations. For general applicability at other telescopes, the results are given in the elevation of the target and its corresponding amount of field rotation. In case of high latitudes, the results shown here are optimistic because of a much smaller amount of field rotation (see Appendix \ref{app:field_rotation}). However, at these high latitudes no major direct imaging facility exists. The model consists of 60 images (spaced over 1 hour of observations), 400$\times$400 pixels each, see Figure \ref{fig:model_layers}. These reductions can be seen in Figure \ref{fig:flux_matrices}, in which the first and second row show the amount of recovered flux and numerical noise after 500 iterations in percentages of the total flux of the original model for cIADI and PCA-IADI, respectively. The third column is the difference between the two techniques.

As presented in the recovered flux panels in Figure \ref{fig:flux_matrices}, higher elevations and inclinations (more asymmetrical disks) are favourable for the recovery of the disk. The elevation of the disk determines both the total amount of field rotation and the speed at which the rotation occurs (see the vertical axis on the right of Fig. \ref{fig:flux_matrices}). The larger the elevation of the object, the more it will rotate during the observation if observed during its meridian passage. When the elevation is close $90\degree$ however, this rotation is almost instantaneous. Hence, even though the field rotation is maximal, the rotation in that case only happens in between two images only: images in the first half (respectively second half) are almost identical to each other, reducing the amount of flux recovered from the disk. In general, independent of the telescope latitude, the optimal amount of field rotation, and rotation speed, occurs at an elevation of $85\degree$, or a $\sim5\degree$ difference between the declination and latitude. The inclination (or similarly any asymmetry in the disk) affects the recovery of the disk as well. For higher inclinations (or more asymmetrical disks), when subtracting the median of the data set from the images, less self-subtraction will occur. For this reason, higher inclined disks are recovered better. Both the elevation and inclination combined give rise to the diagonal gradient going from the bottom left to the top right as seen in the panels of Figure \ref{fig:flux_matrices}.

In general, more of the flux is recovered with PCA-IADI than with cIADI. The recovered flux is on average $25\%$ higher for PCA-IADI than for cIADI. As the panel in Figure \ref{fig:flux_matrices} with the differences in recovered flux between cIADI and PCA-IADI shows, PCA-IADI mainly improves upon the recovery of the disk with cIADI at high inclinations and low elevations or little field rotation. cIADI is more affected by the small differences between the images due to the low elevation, causing less flux to be recovered because more of the disk is self-subtracted. On the other hand, for PCA-IADI the small evenly spaced rotations between every image is still enough for it to not remove the disk completely, resulting in a better recovered disk than with cIADI. Going to higher elevations, the difference in recovered flux is less apparent. At these high elevations, more and faster field rotation occurs, giving rise to a data set essentially containing two similar sets of images, which is easier for PCA to fit to and more of the disk will be subtracted. Hence, in practice, it would be best to apply PCA-IADI to data sets with images with relatively constant rotation between images, while cIADI can also be applied to data sets containing less evenly rotated images.

The noise generated while using the pipeline, especially when the stellar speckle halo is present, is an important factor in deciding which reduction technique to use. As the second column in Figure \ref{fig:flux_matrices} shows, the more the images need to be rotated, and the higher the asymmetry is in the disk, the more numerical noise is generated. While the latter is not as detrimental when using real observations, speckle noise will have the same effect, which will be discussed in successive sections. The introduction of numerical noise at these higher elevations and inclinations is originating from the interpolation while rotating the disk and sharp edges in the model due to the high inclination. 

\begin{figure*}[t!]
    \centering
    \includegraphics[width=\textwidth]{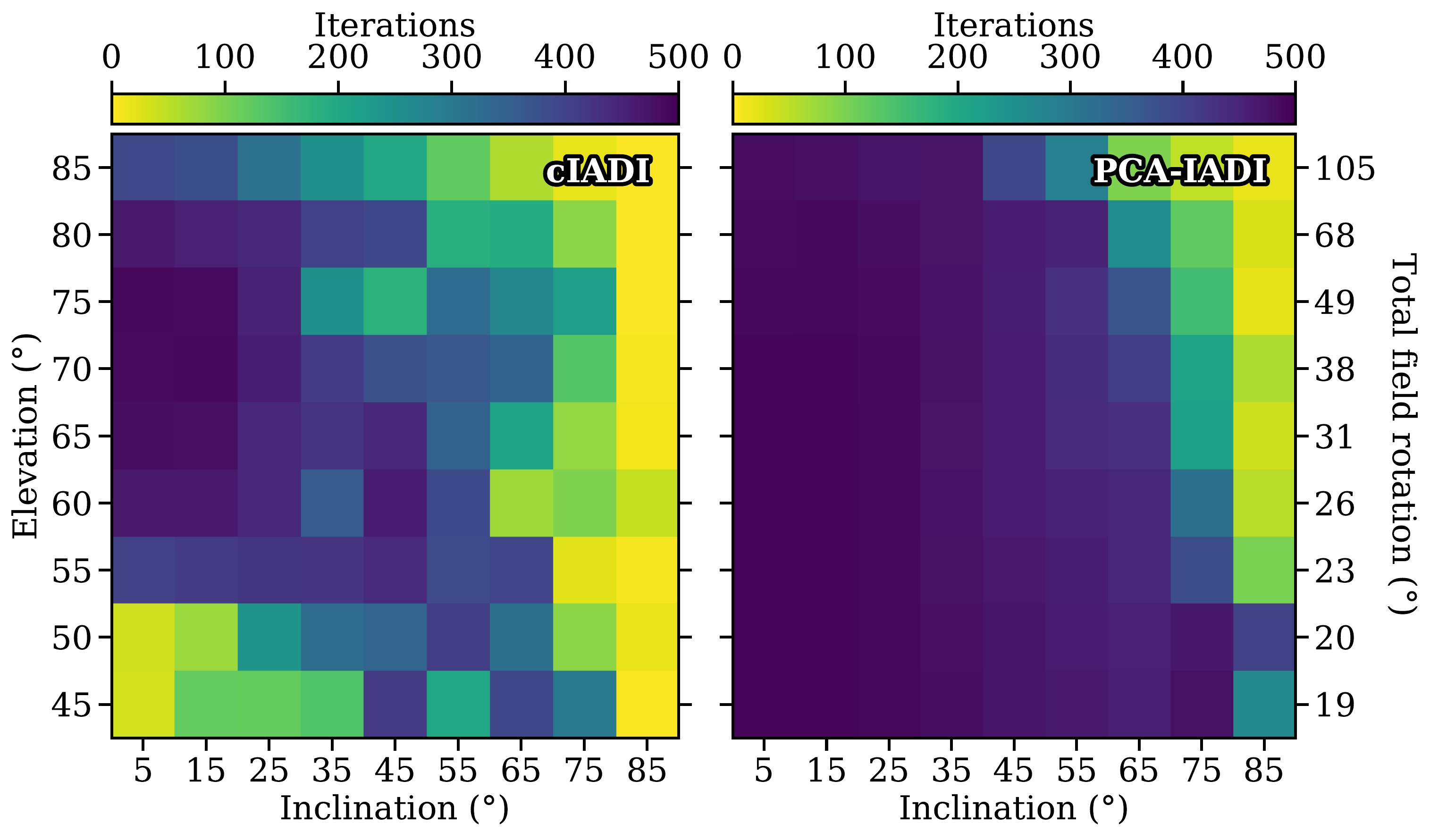}
    \caption{The number of iterations necessary to reach 99\% of the flux recovered after 500 iterations, i.e. the convergence speed.}
    \label{fig:convergience_plot}
\end{figure*}

PCA-IADI generates more noise than cIADI (see Figure \ref{fig:flux_matrices}). On average the amount of numerical noise is $3.3\%$ higher with a maximum difference of $22.0\%$. This is especially a problem when iterating too many times over the data. Because PCA-IADI is also fitting to the numerical noise generated, this noise is amplified over time. Consequently, one needs to know what amount of iterations is at least necessary to recover the disk. This is inferred from Figure \ref{fig:convergience_plot}. This figure shows the number of iterations it took to recover $99\%$ of the amount of flux recovered after 500 iterations for cIADI (left) and PCA-IADI (right). In other words, it indicates how fast the models converge to a particular value. Figure \ref{fig:convergience_plot} shows that, while more asymmetries in the disk (due to inclination) generates more numerical noise in the result, no more than 100 iterations are necessary to recover the disk. Consequently, the amount of numerical noise can be decreased by using less iterations for these highly asymmetric disks. For both cIADI and PCA-IADI, more than 500 iterations are necessary to recover the flux at intermediate elevations and low inclinations. Especially PCA-IADI is slow in converging to a final result. This is likely due to PCA-IADI first subtracting the mean of the data set from each image before fitting the principal components, which are subtracted from the data set as well. This much more aggressively removes the common modes in the images of the data set than cIADI does, which results in a much slower convergence. For cIADI, the final amount of recovered flux in the bottom left part of the plot is already quite low (see Figure \ref{fig:flux_matrices}). The number of iterations to reach this low amount of flux is already reached after about 100 iterations, which explains the relatively fast converging speed at the bottom left of the plot for cIADI. While PCA-IADI does recover more, the high number of iterations necessary makes it more worthwhile to use cIADI for these low elevations (small amount of field rotation), such that the amount of noise introduced due to the iteration process is kept at a minimum.

To conclude, the full disk can be recovered using both versions of IADI, depending on the elevation (or field rotation) and inclination (or asymmetry) of the disk. Due to the slow converging speed of PCA-IADI, cIADI is better suitable for more symmetrical disks. PCA-IADI on the other hand favours highly asymmetrical disks. Due to differences in spacing between images depending on the elevation, cIADI should be used in datasets with large differences in rotation between images, while PCA-IADI works best for small rotational differences. Lastly, PCA-IADI introduces the most noise to the reduction, and amplifies this. Hence, only limited number of iterations should be done with it.


\begin{figure*}[t]
    \centering
    \includegraphics[width=0.97\textwidth]{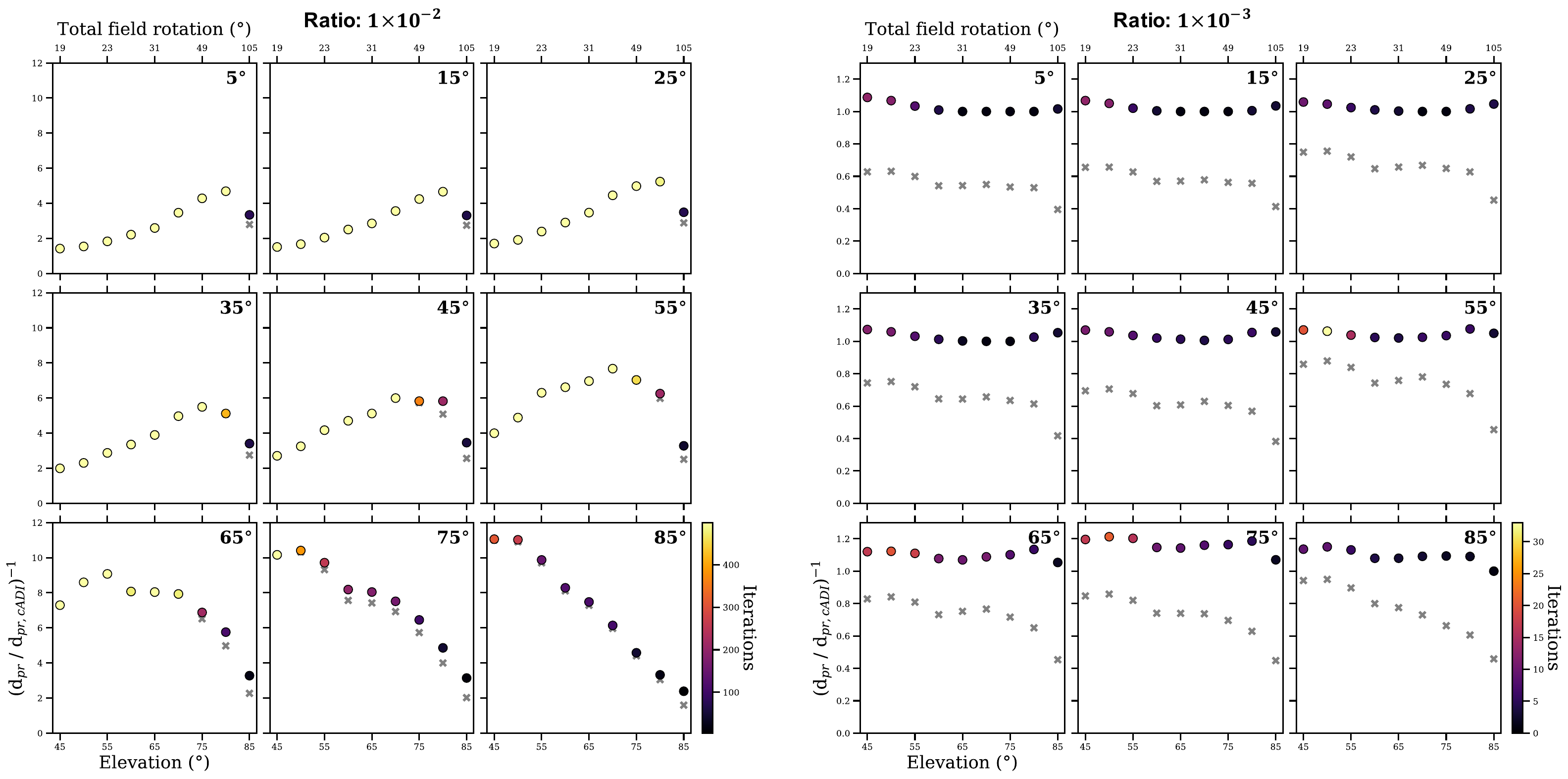}
    \caption{The Procrustes distance of cIADI normalized by the Procrustes distance obtained from cADI of the retrieved phase functions for different inclinations and elevations/total field rotation. The circles show the maximum value out of 500 iterations. The final values are shown as a crosses. Note that the right panel has a different color scale than the left panel.}
    \label{fig:proc_grid_cIADI}
\end{figure*}

\begin{figure*}[h]
    \centering
    \includegraphics[width=0.97\textwidth]{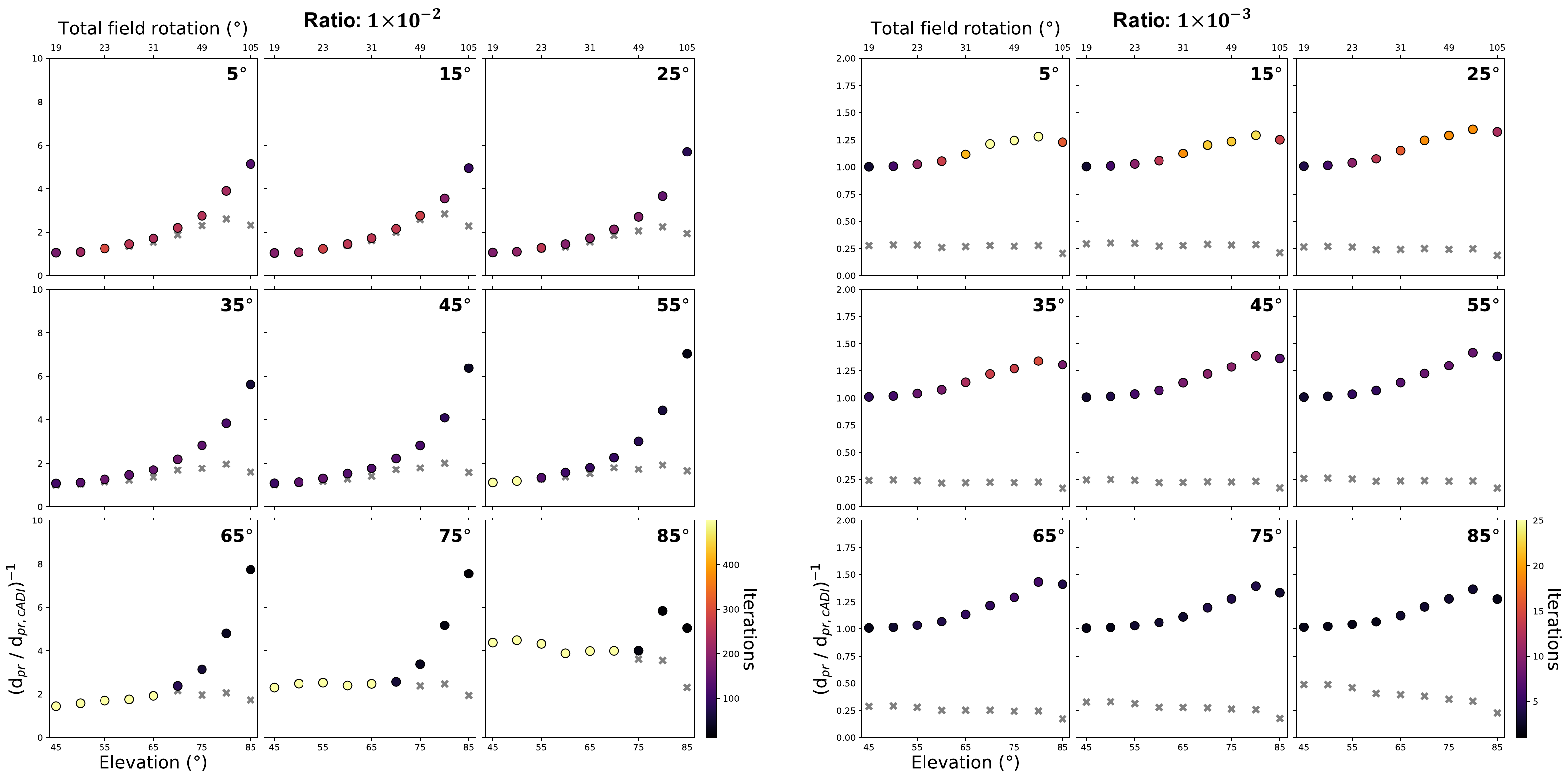}
    \caption{The Procrustes distance of PCA-IADI with 5 principal components normalized by the Procrustes distance obtained from cADI of the retrieved phase functions for different inclinations and elevations/total field rotation. The circles show the maximum value out of 500 iterations. The final values are shown as a crosses. Note that the right panel has a different color scale than the left panel.}
    \label{fig:proc_grid_PCA}
\end{figure*}

\subsection{Test on model inserted into star}
\label{sec:model_and_star}



\begin{figure*}[b]
    \centering
    \includegraphics[width=\textwidth]{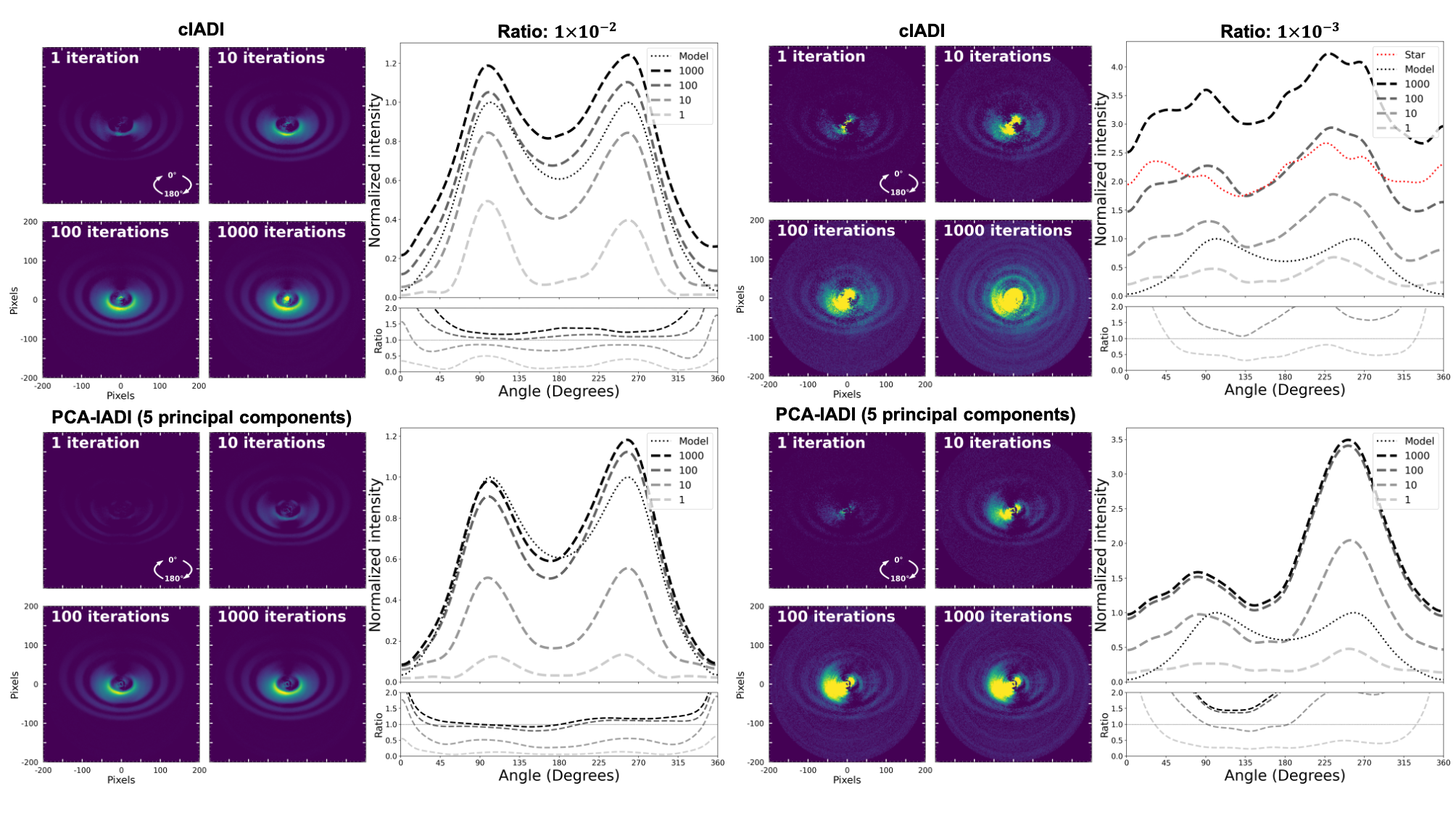}
    \caption{\textbf{Top row:} cIADI reduction of the disk model inserted into a star with a ratio of $1\times10^{-2}$ (left) and $1\times10^{-3}$ (right). The red dotted line is a reduction of the data set without the model inserted. \textbf{Bottom row:} PCA-IADI reduction with 5 principal components of the disk model inserted into a star with a ratio of $1\times10^{-2}$ (left) and $1\times10^{-3}$ (right).}
    \label{fig:disk_and_star}
\end{figure*}

The model was inserted into observations made by SPHERE/IRDIS on the VLT of the star GSC~08047-00232. Both the SED of this system as well as the SPHERE polarimetric observations show that no disk resides around this star. In addition, the data were taken in the same observation mode as a majority of new disk observations (i.e., pupil stabilized observations in polarimetric mode with the BB\_H filter), so the instrument characteristics should be similar to most recent observations. This observation consisted of a total of 40 images instead of the previously used 60 images, but is still simulated with a one hour observation time around meridian. The model and the star are combined by normalizing the model and the data by dividing by their respective maximum values and then added together after multiplying them by a ratio factor of either $1\times10^{-2}$ or $1\times10^{-3}$. For this section, four grids of reductions are done, see Figures \ref{fig:proc_grid_cIADI} and \ref{fig:proc_grid_PCA} for cIADI and PCA-IADI respectively. In addition, one specific case is shown: a model with an elevation of $85\degree$ and an inclination of $45\degree$ for both cIADI and PCA-IADI in Figure \ref{fig:disk_and_star}.

The normalized Procrustes distance on the vertical axes in Figures \ref{fig:proc_grid_cIADI} and \ref{fig:proc_grid_PCA} can be interpreted as the factor by which the recovered phase function improved (or goodness-of-recovery) with respect to classical angular differential imaging (either via using the median or using \texttt{PynPoint} for respectively cIADI and PCA-IADI). For both PCA-IADI and cIADI, the goodness-of-recovery clearly depends on the inclination and elevation of the object in the case of a ratio of $1\times10^{-2}$. In agreement with the previous section, for higher inclinations the model reaches the highest possible improvement factor faster than for low inclined disks due to the increase in asymmetry. In general, cIADI can be run for larger number of iterations than PCA-IADI in high signal to noise cases, while also improving the phase function recovery up to factors of 11. 

When there are too many iterations, the recovered phase function goes above the intensity of the model phase function and may change shape, decreasing the goodness-of-recovery, as indicated by the crosses, especially for the high elevation and inclination cases. This can also be seen in the reduction for a $45\degree$ inclination $85\degree$ elevation model for both a ratio of $1\times10^{-2}$ and $1\times10^{-3}$ shown in Figure \ref{fig:disk_and_star}. For high signal to noise, the original phase function shape is well recovered after only 10 iterations (left panel Figure \ref{fig:disk_and_star}), with 100 iterations the best fitting phase function compared to the original model. Larger number of iterations are starting to reconstruct the star, which can be seen as signal in the center of the image (see Figure \ref{fig:disk_and_star}). Especially in the low signal to noise case, the reconstructed star is dominating the recovered phase function. Additionally, due to the iterative process, the reconstructed star adds circular shapes to the image, making it more difficult to differentiate between the disk and the star. The iteration process also amplifies these values due to interpolation and generates a bright spot in the middle of the image after 1000 iterations. Hence, an increase in Procrustes distance (and thus a decrease in Figures \ref{fig:proc_grid_cIADI} and \ref{fig:proc_grid_PCA}) can be seen. PCA-IADI suppresses this effect, but also recovers less of the signal, only improving the recovery of the phase function up to factors of 8.

\begin{figure*}[b]
    \centering
    \includegraphics[width=\textwidth]{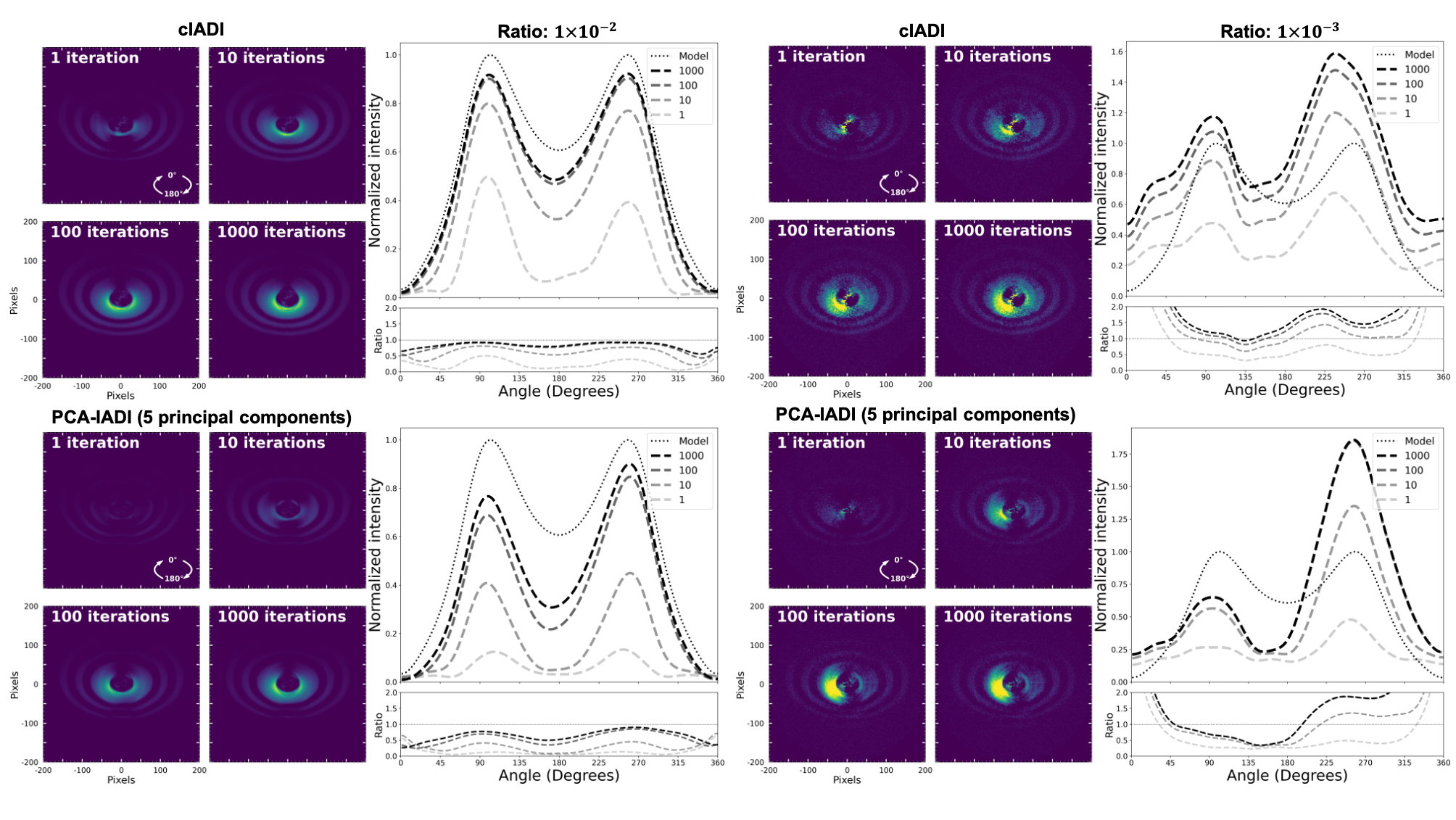}
    \caption{Same reductions as in Figure \ref{fig:disk_and_star}, but now with a mask and threshold which limit the feedback of stellar signal.}
    \label{fig:disk_and_star_threshold}
\end{figure*}

In the top right panel in Figure \ref{fig:disk_and_star}, for a ratio of $1\times10^{-3}$, the reconstruction of the stellar PSF is much better visible (see the red dotted line, which is a reduction of the data set without the model inserted). This much larger impact of the reconstruction of the star also impacts the values of the Procrustes distance (see Figures \ref{fig:proc_grid_cIADI} and \ref{fig:proc_grid_PCA}). Most of the reductions come to the minimum Procrustes distance after only a few tens of iterations, and in general the recovery only improved by a factor up to 1.2, after which the construction of the star dominates the recovery. The stellar PSF also introduces an asymmetry to the phase function visible in both the cIADI and the PCA-IADI reductions at around 90$\degree$ and 270$\degree$. The noise is amplified to higher values than the value in the original model. As long as no more than a few tens of iterations are done, the disk does become better visible (see the top right panel of Figure \ref{fig:disk_and_star}). Especially at larger distances, where the star affects the recovery less, the rings of the model do become better visible. Like the high signal to noise case, PCA-IADI is much more capable of suppressing the reconstruction of the star. PCA-IADI works better in the low signal to noise case, improving the recovery by factors up to 1.5, compared to 1.2 for cIADI.

To conclude, cIADI should be mainly used for the high signal to noise cases, where the number of iterations does not have much effect on the recovery of the disk, except for the high elevation (high amount of field rotation) and high inclination (very asymmetrical disks) cases. PCA-IADI on the other hand is much more aggressive in removing the stellar PSF and thus should mainly be used in the low signal to noise cases. However, one should be careful with PCA-IADI, more than 10s of iterations introduce ring-like artefacts and negatively impacts the overall recovery of the disk.

\subsection{Mask and threshold}
\label{sec:mask_and_threshold}
The previous section demonstrates that the introduction of a star might be detrimental to the disk recovery. The effects of the star are increasing with decreasing SNR of the disk signal. The different effects that were discussed can all be traced back to the same issue, i.e. that the process was not discriminating between signal and noise for the feedback loop. In this section, the feedback of signal will be controlled via a mask and a threshold.

A binary mask can be used to indicate where the disk resides, such that only signal of where the disk should be is fed back. The mask is applied after the images have been rotated back to their original position to be subtracted from the original data set, during step 6 in Figure \ref{fig:IADI}. Everything not below the mask is set to zero, such that this is not subtracted from the original set of images. While a mask can be easily generated from the model, for real observations, if available, polarimetric data of the disk can be used to generate a binary mask. However, we should note that the complete disk is not necessarily very polarized \citep[see e.g.,][]{Ginski2021}. Especially for more inclined disks low scattering angles might be excluded. So, as should be customary, a mask should be used with caution.

Before this section, all signal above zero was fed back. This includes noise, which results in generating false positive signal and amplifying noise, which also happens within the mask. A threshold will minimize this feedback and can be implemented in addition to the mask. This threshold is dynamic, after every iteration the threshold is calculated again. Furthermore, it depends on the distance from the star because of the star being bright at the center of the image and reducing in brightness going to the edges. Moreover, there will be more photon noise and residual speckle noise closer to the star. Hence, the standard deviation changes significantly going outward. So a different standard deviation, on which the threshold is based, is used depending on its separation from the center. This should suppress the feedback of noise even more and hence the generation of false positive disk signal. The threshold system is implemented by dividing the disk into multiple annuli, for each of which the standard deviation of the pixel values is calculated during every iteration. For each annulus, the pixels below the threshold times the standard deviation of that annulus are set to zero. For instance, for a threshold of 0.5, every value below 0.5 times the standard deviation in that annulus is set to zero. 

\begin{figure*}[b!]
    \centering
    \includegraphics[width=\textwidth]{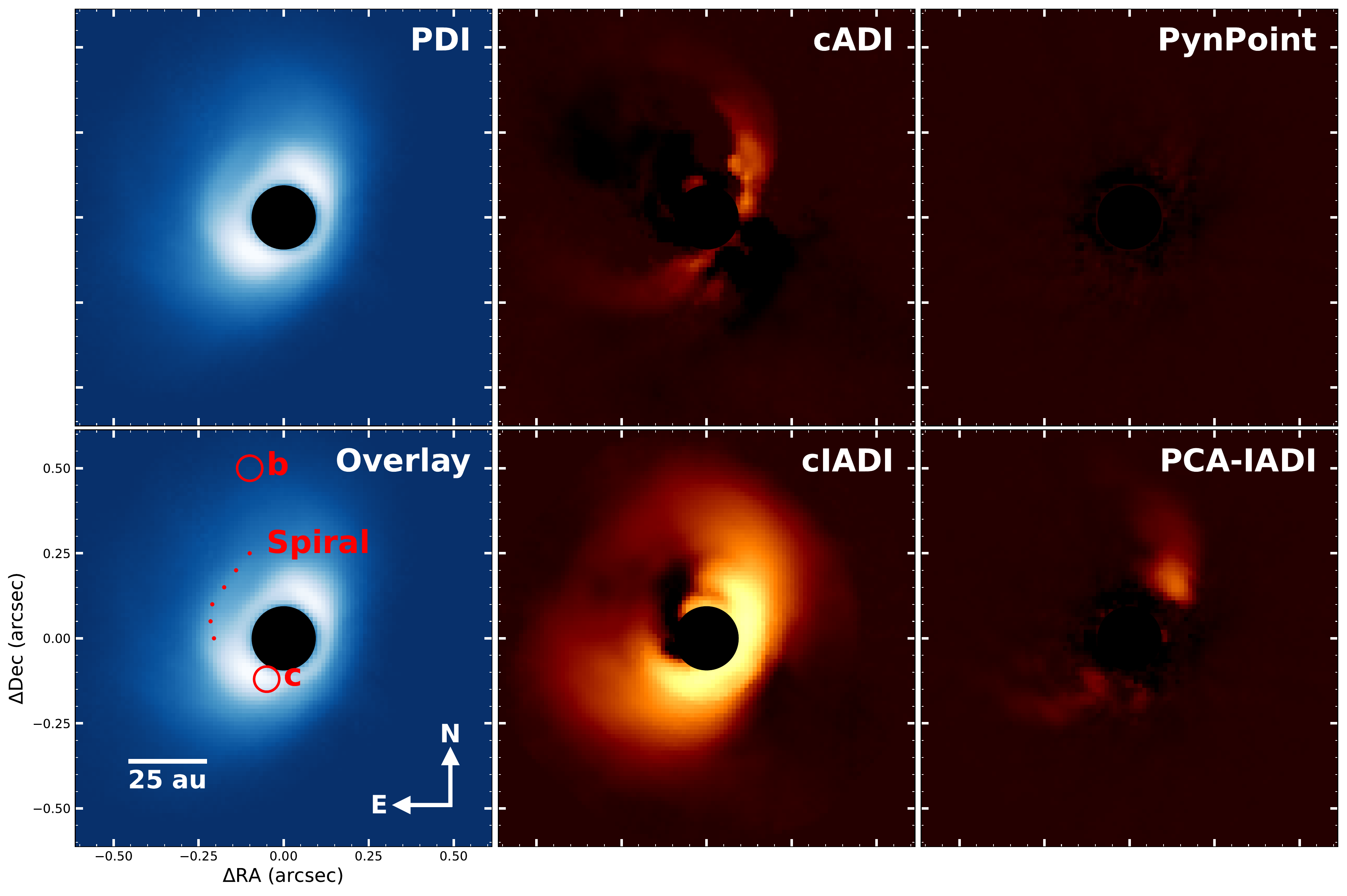}
    \caption{HD~100546 reduced with different reduction techniques. The overlay shows the positions of the planet candidates \citep{Currie2015} and the spiral arm \citep{Avenhaus2014}.}
    \label{fig:HD100546_overview_plot}
\end{figure*}

A reduction with a threshold and a mask can be found in Fig. \ref{fig:disk_and_star_threshold}. The main difference in the left two panels of Fig. \ref{fig:disk_and_star_threshold} compared to Fig. \ref{fig:disk_and_star} is that in the center of the image the star is not fed back. This reduces the amplification of the signal in the center of the image, resulting in that the recovered phase function did not surpass the intensity of the phase function of the input model. The Procrustes distances of the mask and dynamic threshold reduction are lower than the reductions without. The Procrustes distances is reduced from 0.136 for the star and disk only reduction to 0.110 with the mask and dynamic threshold implemented. This is a factor of 1.2 improvement.

Going to the $1\times10^{-3}$ ratio reduction in the left two panels of Figure \ref{fig:disk_and_star_threshold}, the phase function is dominated by the inner ring of the disk, a large part of which has stellar signal. When using a mask, still signal of the star is fed back, which is one of the basic problems of a mask. The threshold may help in reducing this feedback of stellar signal, but it may not reduce it completely. The Procrustes distances show that again the stellar signal is being reconstructed, after 10 iterations the Procrustes distances are increasing again (from 0.180 to 0.769 for cIADI and from 1.68 to 2.14 for the 5 principal component PCA-IADI reduction). The reduction with no mask or dynamic threshold show a worse recovery than the reduction with a mask and a dynamic threshold. Lastly, the side facing the observer is clearly brighter than the other side of the disk, which originates from the original `pseudo'-phase function of the model.

The introduction of a mask and a dynamic threshold did reduce the creation of artefacts significantly. In all reductions with PCA-IADI for different principal components for a ratio of $1\times10^{-2}$ no bright rings can be seen anymore. The recovered images are clearly a better representation of the original model. However, the Procrustes distances computed are higher than without the dynamic threshold and the mask (1.68 compared to 1.07). This is mainly due to less flux being recovered around 180$\degree$, where the recovered phase function has a deeper valley than the original model. However, the signal mainly consists of disk signal instead of being dominated by the amplified noise, which is an improvement. Moreover, where first the recovered phase functions even exceeded the original model, now the dynamic threshold and mask ensure that this does not happen. Hence the following is concluded: the mask and the dynamical threshold should not be used for low SNR cases, in the case of high SNR they can help in recovering the disk, though should be used with care.

\begin{figure*}[b!]
    \centering
    \includegraphics[width=\textwidth]{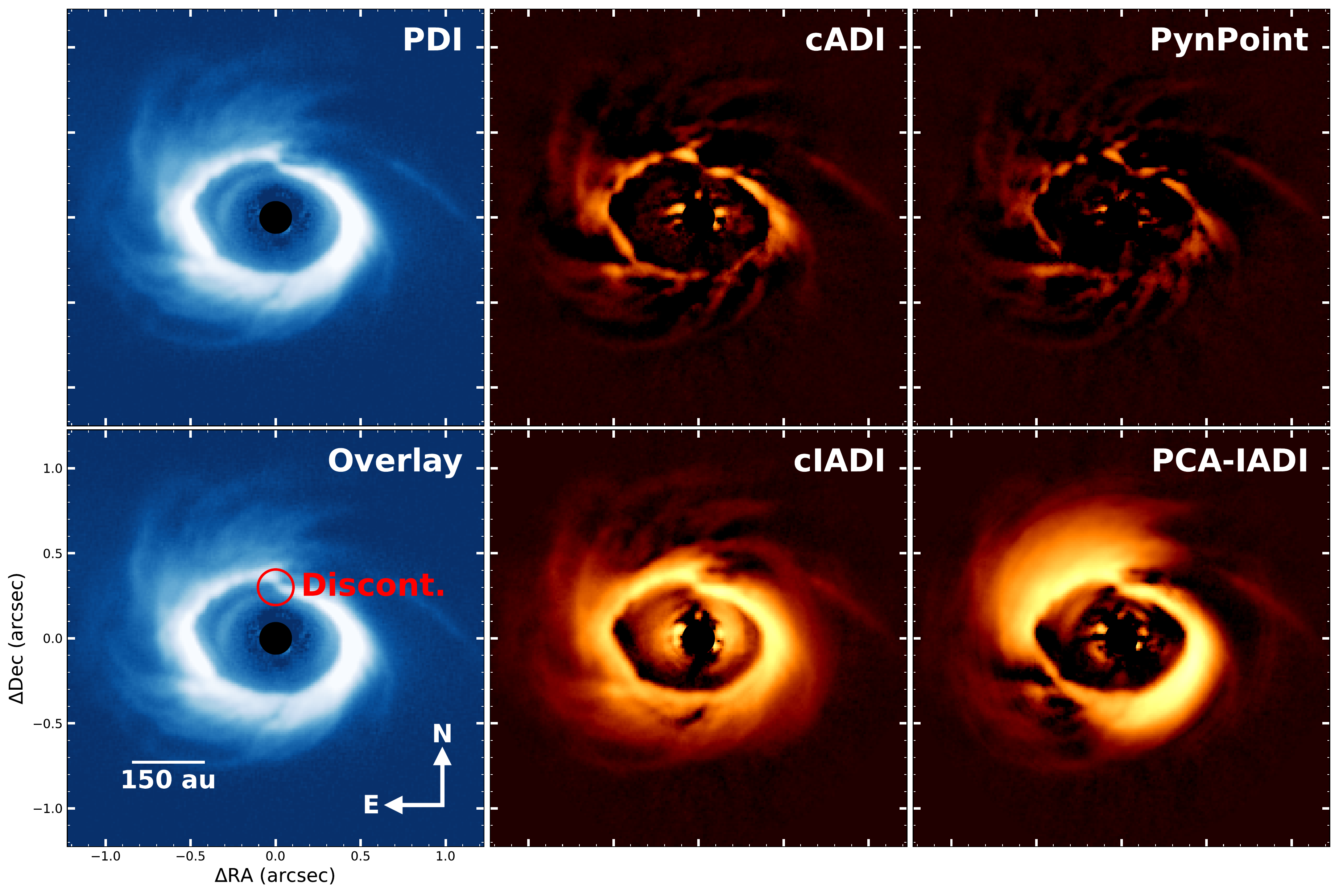}
    \caption{HD~34700~A reduced with different reduction techniques. The overlay shows the discontinuity as found by \citet{Monnier2019}.}
    \label{fig:HD34700_overview_plot}
\end{figure*}

\section{Application to real data}
\label{sec:application_to_data}

\subsection{HD~100546}

In the disk of HD~100546 spiral structures have been found \citep{Avenhaus2014,Follette2017, Sissa2018, pineda2019} and the presence of two proto-planet candidates was proposed (\citealt{quanz2013, quanz2015, Currie2014, Currie2015}), the status of which however remains controversial (\citealt{Garufi2016,Rameau2017}). In Figure \ref{fig:HD100546_overview_plot}, we show polarimetric and full intensity scattered light SPHERE data taken on the 18th of February 2019. This data were taken 30 minutes after meridian passing and had a duration of 2 hours resulting in 127 exposures with 34.3$\degree$ of field rotation. The average field rotation between each exposure is 0.096$\degree$ with a standard deviation of 0.013$\degree$, not including a large difference of 22.3$\degree$ between exposures 63 and 64. This small difference between each image is due to the low elevation of the source.

The reductions of the data set are shown in Figure \ref{fig:HD100546_overview_plot}. The polarized differential imaging result, produced with the IRDAP pipeline (\citealt{vanHolstein2017,vanHolstein2020}), is shown on the top left. The classical ADI pipeline reduction is shown in the top center image. The reduction with \texttt{\texttt{PynPoint}} \citep{Stolker2019} for 1 principal component is shown in the top right image. For more components, the disk was completely removed. The relatively low asymmetry of the disk in addition to the moderate amount of field rotation, Fig. \ref{fig:convergience_plot} shows that around 300-400 iterations can be done to maximize the amount of flux recovered. While the disk is visible in the raw data, the ratio between the disk and the star is likely a bit worse than shown in the left panel of Figs. \ref{fig:proc_grid_cIADI} and \ref{fig:proc_grid_PCA}. Thus, $\sim$300 iterations will be suitable for this dataset. In addition, the data set was processed with a mask made with the PDI image with a threshold of 20 (i.e. every pixel with a value above 20 in the PDI image is fed back) and a dynamic threshold of 0.01. The standard PCA-IADI pipeline with 1 principal component was used for the corresponding result, without adding a mask or dynamic threshold. The before mentioned positions of the spiral and the planet candidates are shown in the overlay.

The classical ADI result in the top center of the image shows two structures coming from the center of the image going north-eastward. \citet{Follette2017} and others reported these features as possible spiral structures in the disk itself, with planet b positioned at the end of the top arm. However, as \citet{Garufi2016} already showed in their reductions, these are artifacts of the reduction method. Mainly the edges of the disk are recovered, and indeed  with cIADI, more of the disk is recovered and none of these structures are present anymore. The recovered full intensity disk is of similar size as the disk recovered with PDI. The spiral reported by \citet{Avenhaus2014} seems to be present in the cIADI recovered image as well. No signal of any planet can be seen in the disk. The `dark lane', or sharp drop in brightness, in the south-west part of the disk first reported by \citet{Avenhaus2014} in polarized intensity, is present here in full intensity as well.

The \texttt{PynPoint} reduction with 1 principal component shows a faintly visible disk edge at the same position as the arms in the cADI reduction. PCA-IADI does not recover as much of the disk as cIADI does. Only parts of the edges of the disk are visible. This stark difference between cIADI and PCA-IADI arises from both the amount and speed of field rotation in the data, which is only 34.3$\degree$ (65\% of which happens between two observations) and constant between each image. As was discussed in Section \S\ref{sec:model_only}, PCA-IADI works best for constant field rotation with no large gaps in field rotation. On the other hand, cIADI works best for large differences in field rotation. Hence, while cIADI can recover large parts of the disk, PCA-IADI only recovers the part of the disk which changed the most during the observation.

\begin{figure*}[b!]
    \centering
    \includegraphics[width=\textwidth]{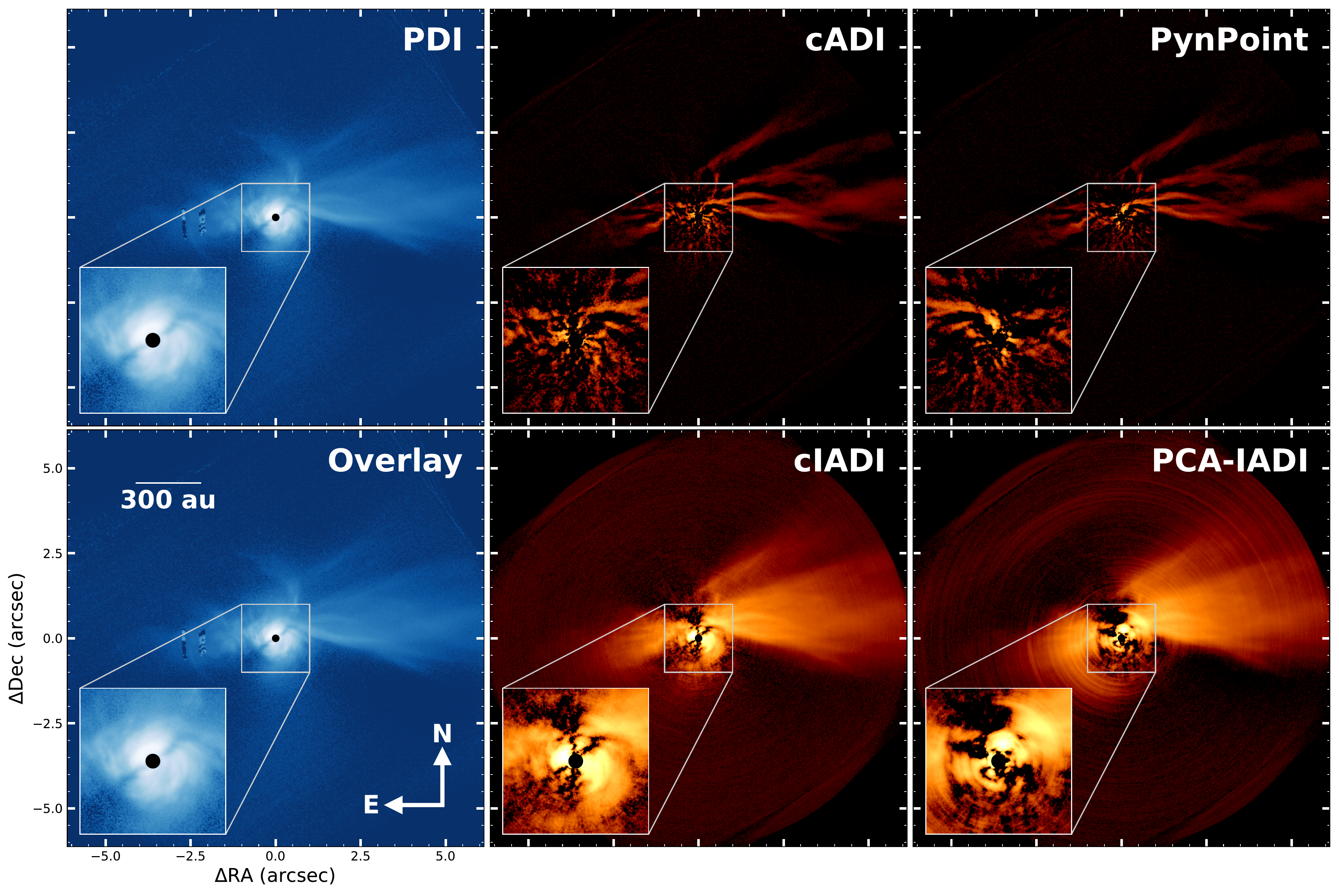}
    \caption{SU~Aurigae reduced with different reduction techniques.}
    \label{fig:SUAur_overview_plot}
\end{figure*}

\subsection{HD~34700~A}
The HD~34700~A disk consists of a large circumstellar dust ring with a major-axis of 175 au \citep{Monnier2019} to which multiple spiral arms have found to be attached to, possibly up to 8 in total \citep{Monnier2019}. A discontinuity can be seen in the northern part of the dust ring as well. In Figure \ref{fig:HD34700_overview_plot}, we show polarimetric and full intensity scattered light SPHERE data taken on the 27th of October 2019. This data were taken centered on the meridian with a duration of 1.5 hours. It consists of 64 exposures with 35.3$\degree$ of field rotation. Each image is on average rotated by 0.56$\degree$ due to field rotation. The disk is very asymmetric due to the many spirals present, which translates to a relatively high inclination in Figs. \ref{fig:convergience_plot}, \ref{fig:proc_grid_cIADI} and \ref{fig:proc_grid_PCA}. In combination with the elevation (and amount of field rotation) the optimal number of iterations is somewhere between 100-200. Lastly, in the raw data the disk is very well visible. The disk is relatively bright compared to the star, which makes it similar to the ratio $1\times10^{-2}$ case discussed in Sections \ref{sec:model_and_star} and \ref{sec:mask_and_threshold}. For these reductions, the optimal number of iterations are indeed around 200 iterations. The reductions of the data set are shown in Figure \ref{fig:HD34700_overview_plot}. The polarized differential imaging result is shown on the top left. The classical ADI pipeline reduction is shown in the top center image. The reduction with \texttt{PynPoint} for 1 principal component is shown in the top right image, which gave the best result compared to different numbers of principal components.

The data set was processed with cIADI for 200 iterations, together with a mask, made with the PDI image with a threshold of 10, and a dynamic threshold of 0.01. The PCA-IADI pipeline with 1 principal component was used with the same mask and dynamic threshold as cIADI.

In Figure \ref{fig:HD34700_overview_plot} the position of a discontinuity is shown in the overlay. In the PDI image all of the spiral arms observed by \citet{Monnier2019} can be seen. One large spiral arm on the west side of the disk going south-westward, multiple spiral arms on the north and north-east side of the disk and on the south-east part of the disk. With cADI and \texttt{PynPoint} the outlines of these spirals and the edges of the disk can be seen. The large extending arms are recovered well already with the classical reduction techniques due to their asymmetric nature. The discontinuity can be seen in these reduction as well. However, large parts of the dust ring are not recovered due to self-subtraction. With cIADI the parts with the missing flux in the south of the disk are recovered. With PCA-IADI on the other hand, mainly the north of the disk is recovered. For both cases the recovered total intensity image of the bright ring compares well with the total intensity reduction using reference differential imaging presented in \citet{Uyama2020} for Subaru/CHARIS data.

\subsection{SU~Aurigae}
The main feature of SU~Aur is a large tail \citep{chakraborty2004, Jeffers2014, deLeon2015} extending outwards to up to 1000 au on the west side of the disk. While multiple explanations of this arm have been proposed such as a cloud remnant \citep{Jeffers2014} or a collision with a stellar intruder \citep{Akiyama2019}, recent work by \citep{Ginski2021} shows that, by using results reported here, the tails (both the large tail in the west and the shorter one in the north) originate from still infalling material. In Figure \ref{fig:SUAur_overview_plot}, we show the polarimetric and full intensity scattered light SPHERE data taken on the 14th of December 2019. The data were taken 15 minutes after meridian passing and took around 1 hour of observation time. The data consists of 104 exposures of 32 seconds with a total of only 18.1$\degree$ of field rotation. On average 0.18$\degree$ of rotation is in between each image. The large tail of the disk is visible in the raw data at large separations, but the disk itself is not visible.

The reductions of the data set are shown in Figure \ref{fig:SUAur_overview_plot}. The polarized differential imaging result is shown on the top left. The classical ADI pipeline reduction is shown in the top center image. The reduction with \texttt{PynPoint} for 2 principal components is shown in the top right image, which gave the best result compared to different numbers of principal components. The disk is very asymmetric, hence, not too many iterations are necessary, especially given that more noise will be produced as well (see Fig. \ref{fig:flux_matrices}). Taking the results in Figs. \ref{fig:convergience_plot}, \ref{fig:proc_grid_cIADI} and \ref{fig:proc_grid_PCA} into account, the optimal number of iterations is therefore around 100 iterations. No masking has been used. The standard PCA-IADI pipeline was used with 2 principal components.

In the PDI image, the east and westward going tails are well visible, the inset also shows the inner disk. Classical ADI only recovers the edges of the tails. The tail is the most asymmetric part of the disk and will overlap the least in the data set, so it will also be recovered first. However, due to the low amount of field rotation and the small differences in between each image, almost nothing of the disk in the center of the image is recovered. cIADI does not help in recovering the disk itself as well. The structure which can be seen is likely due to the reconstruction of the star. They appear too circular to be the disk. However, the large extended tails in both the east and west direction are very well recovered. More of the structure which is also visible in the PDI result is now visible in the full intensity image as well. No significant numerical rings are produced during the iteration process even though no mask or dynamic threshold are used in this reduction.

With \texttt{PynPoint} the extended tails, which were also recovered with cADI, can be seen here. Compared to cADI the residual pattern changes but the disk is still not recovered. PCA-IADI recovers more of the tail. Just like with cIADI, both the eastern and western tails are filled in due to the iteration process. The tail in the northern part of the westward going tail is much brighter than with PDI. However, due to iteration process, the numerical rings are amplified by the PCA-IADI reduction (as was expected from Sections \ref{sec:model_only} and \ref{sec:model_and_star}). Hence, cIADI would be better suited for this disk.

\section{Summary}
\label{sec:summary}
In Section \ref{sec:model_only} the effectiveness and efficiency of IADI is assessed for two different algorithm versions: Classical Iterative Angular Differential Imaging (cIADI), which determines the PSF of the star via taking the median of the data set, and Principal Component Analysis Iterative Differential Imaging (PCA-IADI), which determines the PSF of the star via Principal Component Analysis (PCA, identical to the GreeDS algorithm presented by \citealt{Pairet2021}). Processing is performed for model-only and star+model cases. Here follows a short summary containing our main conclusions.


Asymmetries are an important factor in recovering the disk. These asymmetries can either be caused by the inherent structure of the disk, or by the inclination of the disk. Due to less of the disk overlapping in the data set, more of the disk is recovered. The other important factor in recovering the disk is the amount of field rotation caused by the declination of the disk and latitude of the telescope, i.e. the elevation of the disk. In general, the more field rotation, the more of the disk is recovered, due to less overlap between images. However, how fast the field rotation occurs is important as well. The higher the elevation, the faster the rotation happens. Consequently, most of the images in the data set will look similar because most of the rotation happens only between a few of the images in the data set. Which is well-established behavior of any ADI algorithm. For IADI on the other hand, different behavior is found for different speeds of field rotation. cIADI works best for data sets containing high amounts of field rotation, while PCA-IADI works better with data sets containing evenly spaced rotated images. When the images are evenly spaced rotated, the difference between every single image is maximal which gives PCA-IADI the best opportunity to distinguish the disk from the stellar PSF in the data set.

Keeping these limitations in mind, we find that both variations of IADI significantly increase the recovered disk flux in the model-only processed data sets. For high inclination disks ($i>75\degree$) which pass meridian within $\sim$ 10$\degree$ of the local zenith, the throughput is close to 100\%. Even for the worst case of a nearly face-on disk ($i=5\degree$), a $\sim$20\% throughput is achieved assuming that the target passes meridian within $\sim$ 15$\degree$ of the local zenith. For the Paranal observatory this requirement would be fulfilled for several of the abundant nearby star forming regions such as the Lupus cloud complex or Corona Australis. 

For cIADI the disk is best recovered in the high flux ratio case, but the star still introduces asymmetries into the recovered phase functions. For PCA-IADI, depending on the amount of principal components used, numerical artifacts were made worse or were completely removed. In general the application of PCA-IADI led to reduced throughput but also reduced stellar noise. However, for the low flux ratio case the recovered phase functions were dominated by stellar speckle noise in both techniques. By applying the Procrustes metric we found that, depending on the flux ratio, a small number of iterations ($\sim$10) is optimal, since the stellar speckle halo tends to be dominantly reconstructed at higher iterations. The mask and dynamic threshold do significantly improve the recovery of the disk as well as the phase function, especially in the $1\times10^{-3}$ ratio case for cIADI. For PCA-IADI the masking technique may introduce false positive signal for low flux ratio disks and needs to be applied cautiously. 

For the three exemplary cases of observed disks, a significant improvement is made on the recovery of the disk in full intensity NIR scattered light. While with previously used techniques, such as classical ADI and \texttt{PynPoint}, only the edges of the disks or spiral arms were recovered, IADI increases throughput for the inner disk and specifically low spatial frequency signal. This is also what is found in theoretical part of this work, the model disks were always recovered from the edges inward, filling in the self-subtracted regions. The best example of this is HD~100546, which had only the edges recovered with cADI (and \texttt{PynPoint}), but with cIADI the disk is reconstructed from these edges inward resulting in a more complete recovery of the disk in full intensity scattered light. These edges were recovered previously (e.g. \citealt{Currie2015, Garufi2016, Rameau2017}), and some were interpreted as real features of the disk. Hence, these interpretations need to be reconsidered thoroughly, because some of these might be artifacts from the ADI method used. While none of the disks are very symmetric, extended features such as long spiral arms were recovered in greater detail than for example the central regions of the disk. This can be seen in SU~Aur, which consists of both a symmetric inner disk and a large westward extending arm. While the arm is recovered in similar detail as the result from PDI, the inner region of the disk is not. Another example is HD~34700~A, where the spiral arms visible in PDI can be seen in classical ADI or \texttt{PynPoint}, but not with as much detail. IADI (either classical or PCA based) recovered the spiral arms well due to their asymmetric nature.

One of the main difficulties with IADI is the generation and amplification of noise. In section \ref{sec:model_and_star} this manifested in ring structures both due to the intrinsic noise of the image as well as the signal coming from the star being smeared out into circles due to the interpolation used for rotating the images during the iterative process. PCA-IADI amplified these noise induced ring structures even more by fitting them in the PCA subtraction process. This causes the noise amplification to be worse with PCA-IADI than with cIADI. This does depend on how bright the disk is compared to the star however. For the reductions of real measurements most disks were relatively bright when compared to the star (close to the theoretical $1\times10^{-2}$ flux ratio case). The further away from the star, the less of a problem these artifacts become, this can be seen in the reduction of SU~Aur. Here, the rings are likely generated from interpolation artifacts, while closer in to the star the residual speckle noise causes the ring structures. The PCA-IADI reduction of SU~Aur had more rings than the cIADI reduction. As is found in Section \ref{sec:model_and_star} as well, PCA-IADI amplifies the noise much more than cIADI, which is clear from this reduction as well.

In conclusion, IADI does improve the recovery of the disk in full intensity NIR scattered light significantly compared to the standard ADI algorithms. After post-processing in many cases the disk is as well recovered as with PDI. Many of the features visible with PDI are also visible with IADI. It does still cope with problems however. The amplification and generation of (numerical and/or correlated) noise is still an issue which has to be dealt with. In the mean time, IADI can work very well for disks which are relatively bright and extended.

\begin{acknowledgements}
SPHERE is an instrument designed and built by a consortium consisting of IPAG (Grenoble, France), MPIA (Heidelberg, Germany), LAM (Marseille, France), LESIA (Paris, France), Laboratoire Lagrange (Nice, France), INAF—Osservatorio di Padova (Italy), Observatoire de Genève (Switzerland), ETH Zurich (Switzerland), NOVA (Netherlands), ONERA (France), and ASTRON (The Netherlands) in collaboration with ESO. SPHERE was funded by ESO, with additional contributions from CNRS (France), MPIA (Germany), INAF (Italy), FINES (Switzerland), and NOVA (The Netherlands). SPHERE also received funding from the European Commission Sixth and Seventh Framework Programmes as part of the Optical Infrared Coordination Network for Astronomy (OPTICON)under grant No. RII3-Ct2004-001566 for FP6 (2004-2008), grant No. 226604 for FP7 (2009-2012), and grant No. 312430 for FP7 (2013–2016). This work makes use of the following software: Python version 3.9, astropy \citep{astropy2013, astropy2018}, matplotlib \citep{Hunter2007}, numpy \citep{Harris2020}, opencv \citep{opencv_library}, pynpoint \citep{Stolker2019}, scipy \citep{2020SciPy-NMeth}. Lastly, we thank an anonymous referee for their careful consideration of our work and for their thoughtful comments which significantly improved the manuscript.
\end{acknowledgements}

\bibliographystyle{aa}
\bibliography{references.bib}

\begin{thebibliography}{60}
\expandafter\ifx\csname natexlab\endcsname\relax\def\natexlab#1{#1}\fi

\bibitem[{{Akiyama} {et~al.}(2019){Akiyama}, {Vorobyov}, {Baobabu Liu}, {Dong},
  {de Leon}, {Liu}, \& {Tamura}}]{Akiyama2019}
{Akiyama}, E., {Vorobyov}, E.~I., {Baobabu Liu}, H., {et~al.} 2019,
  \href{http://dx.doi.org/10.3847/1538-3881/ab0ae4}{\color{blue}\aj},
  \href{https://ui.adsabs.harvard.edu/abs/2019AJ....157..165A}{157, 165}

\bibitem[{{Amara} \& {Quanz}(2012)}]{Amara2012}
{Amara}, A. \& {Quanz}, S.~P. 2012,
  \href{http://dx.doi.org/10.1111/j.1365-2966.2012.21918.x}{\color{blue}\mnras},
  \href{https://ui.adsabs.harvard.edu/abs/2012MNRAS.427..948A}{427, 948}

\bibitem[{{Apai} {et~al.}(2004){Apai}, {Pascucci}, {Brandner}, {Henning},
  {Lenzen}, {Potter}, {Lagrange}, \& {Rousset}}]{Apai2004}
{Apai}, D., {Pascucci}, I., {Brandner}, W., {et~al.} 2004,
  \href{http://dx.doi.org/10.1051/0004-6361:20034549}{\color{blue}\aap},
  \href{https://ui.adsabs.harvard.edu/abs/2004A&A...415..671A}{415, 671}

\bibitem[{{Astropy Collaboration} {et~al.}(2018){Astropy Collaboration},
  {Price-Whelan}, {Sip{\H{o}}cz}, {G{\"u}nther}, {Lim}, {Crawford}, {Conseil},
  {Shupe}, {Craig}, {Dencheva}, {Ginsburg}, {Vand erPlas}, {Bradley},
  {P{\'e}rez-Su{\'a}rez}, {de Val-Borro}, {Aldcroft}, {Cruz}, {Robitaille},
  {Tollerud}, {Ardelean}, {Babej}, {Bach}, {Bachetti}, {Bakanov}, {Bamford},
  {Barentsen}, {Barmby}, {Baumbach}, {Berry}, {Biscani}, {Boquien}, {Bostroem},
  {Bouma}, {Brammer}, {Bray}, {Breytenbach}, {Buddelmeijer}, {Burke},
  {Calderone}, {Cano Rodr{\'\i}guez}, {Cara}, {Cardoso}, {Cheedella}, {Copin},
  {Corrales}, {Crichton}, {D'Avella}, {Deil}, {Depagne}, {Dietrich}, {Donath},
  {Droettboom}, {Earl}, {Erben}, {Fabbro}, {Ferreira}, {Finethy}, {Fox},
  {Garrison}, {Gibbons}, {Goldstein}, {Gommers}, {Greco}, {Greenfield},
  {Groener}, {Grollier}, {Hagen}, {Hirst}, {Homeier}, {Horton}, {Hosseinzadeh},
  {Hu}, {Hunkeler}, {Ivezi{\'c}}, {Jain}, {Jenness}, {Kanarek}, {Kendrew},
  {Kern}, {Kerzendorf}, {Khvalko}, {King}, {Kirkby}, {Kulkarni}, {Kumar},
  {Lee}, {Lenz}, {Littlefair}, {Ma}, {Macleod}, {Mastropietro}, {McCully},
  {Montagnac}, {Morris}, {Mueller}, {Mumford}, {Muna}, {Murphy}, {Nelson},
  {Nguyen}, {Ninan}, {N{\"o}the}, {Ogaz}, {Oh}, {Parejko}, {Parley}, {Pascual},
  {Patil}, {Patil}, {Plunkett}, {Prochaska}, {Rastogi}, {Reddy Janga},
  {Sabater}, {Sakurikar}, {Seifert}, {Sherbert}, {Sherwood-Taylor}, {Shih},
  {Sick}, {Silbiger}, {Singanamalla}, {Singer}, {Sladen}, {Sooley},
  {Sornarajah}, {Streicher}, {Teuben}, {Thomas}, {Tremblay}, {Turner},
  {Terr{\'o}n}, {van Kerkwijk}, {de la Vega}, {Watkins}, {Weaver}, {Whitmore},
  {Woillez}, {Zabalza}, \& {Astropy Contributors}}]{astropy2018}
{Astropy Collaboration}, {Price-Whelan}, A.~M., {Sip{\H{o}}cz}, B.~M., {et~al.}
  2018, \href{http://dx.doi.org/10.3847/1538-3881/aabc4f}{\color{blue}\aj},
  \href{https://ui.adsabs.harvard.edu/abs/2018AJ....156..123A}{156, 123}

\bibitem[{{Astropy Collaboration} {et~al.}(2013){Astropy Collaboration},
  {Robitaille}, {Tollerud}, {Greenfield}, {Droettboom}, {Bray}, {Aldcroft},
  {Davis}, {Ginsburg}, {Price-Whelan}, {Kerzendorf}, {Conley}, {Crighton},
  {Barbary}, {Muna}, {Ferguson}, {Grollier}, {Parikh}, {Nair}, {Unther},
  {Deil}, {Woillez}, {Conseil}, {Kramer}, {Turner}, {Singer}, {Fox}, {Weaver},
  {Zabalza}, {Edwards}, {Azalee Bostroem}, {Burke}, {Casey}, {Crawford},
  {Dencheva}, {Ely}, {Jenness}, {Labrie}, {Lim}, {Pierfederici}, {Pontzen},
  {Ptak}, {Refsdal}, {Servillat}, \& {Streicher}}]{astropy2013}
{Astropy Collaboration}, {Robitaille}, T.~P., {Tollerud}, E.~J., {et~al.} 2013,
  \href{http://dx.doi.org/10.1051/0004-6361/201322068}{\color{blue}\aap},
  \href{http://adsabs.harvard.edu/abs/2013A%26A...558A..33A}{558, A33}

\bibitem[{{Avenhaus} {et~al.}(2018){Avenhaus}, {Quanz}, {Garufi}, {Perez},
  {Casassus}, {Pinte}, {Bertrang}, {Caceres}, {Benisty}, \&
  {Dominik}}]{Avenhaus2018}
{Avenhaus}, H., {Quanz}, S.~P., {Garufi}, A., {et~al.} 2018,
  \href{http://dx.doi.org/10.3847/1538-4357/aab846}{\color{blue}\apj},
  \href{https://ui.adsabs.harvard.edu/abs/2018ApJ...863...44A}{863, 44}

\bibitem[{{Avenhaus} {et~al.}(2014){Avenhaus}, {Quanz}, {Meyer}, {Brittain},
  {Carr}, \& {Najita}}]{Avenhaus2014}
{Avenhaus}, H., {Quanz}, S.~P., {Meyer}, M.~R., {et~al.} 2014,
  \href{http://dx.doi.org/10.1088/0004-637X/790/1/56}{\color{blue}\apj},
  \href{https://ui.adsabs.harvard.edu/abs/2014ApJ...790...56A}{790, 56}

\bibitem[{{Avenhaus} {et~al.}(2017){Avenhaus}, {Quanz}, {Schmid}, {Dominik},
  {Stolker}, {Ginski}, {de Boer}, {Szul{\'a}gyi}, {Garufi}, {Zurlo},
  {Hagelberg}, {Benisty}, {Henning}, {M{\'e}nard}, {Meyer}, {Baruffolo},
  {Bazzon}, {Beuzit}, {Costille}, {Dohlen}, {Girard}, {Gisler}, {Kasper},
  {Mouillet}, {Pragt}, {Roelfsema}, {Salasnich}, \& {Sauvage}}]{Avenhaus2017}
{Avenhaus}, H., {Quanz}, S.~P., {Schmid}, H.~M., {et~al.} 2017,
  \href{http://dx.doi.org/10.3847/1538-3881/aa7560}{\color{blue}\aj},
  \href{https://ui.adsabs.harvard.edu/abs/2017AJ....154...33A}{154, 33}

\bibitem[{{Benisty} {et~al.}(2022){Benisty}, {Dominik}, {Follette}, {Garufi},
  {Ginski}, {Hashimoto}, {Keppler}, {Kley}, \& {Monnier}}]{Benisty2022}
{Benisty}, M., {Dominik}, C., {Follette}, K., {et~al.} 2022,
  \href{https://ui.adsabs.harvard.edu/abs/2022arXiv220309991B}{arXiv e-prints,
  arXiv:2203.09991}

\bibitem[{{Benisty} {et~al.}(2018){Benisty}, {Juh{\'a}sz}, {Facchini},
  {Pinilla}, {de Boer}, {P{\'e}rez}, {Keppler}, {Muro-Arena}, {Villenave},
  {Andrews}, {Dominik}, {Dullemond}, {Gallenne}, {Garufi}, {Ginski}, \&
  {Isella}}]{Benisty2018}
{Benisty}, M., {Juh{\'a}sz}, A., {Facchini}, S., {et~al.} 2018,
  \href{http://dx.doi.org/10.1051/0004-6361/201833913}{\color{blue}\aap},
  \href{https://ui.adsabs.harvard.edu/abs/2018A&A...619A.171B}{619, A171}

\bibitem[{{Beuzit} {et~al.}(2019){Beuzit}, {Vigan}, {Mouillet}, {Dohlen},
  {Gratton}, {Boccaletti}, {Sauvage}, {Schmid}, {Langlois}, {Petit},
  {Baruffolo}, {Feldt}, {Milli}, {Wahhaj}, {Abe}, {Anselmi}, {Antichi},
  {Barette}, {Baudrand}, {Baudoz}, {Bazzon}, {Bernardi}, {Blanchard}, {Brast},
  {Bruno}, {Buey}, {Carbillet}, {Carle}, {Cascone}, {Chapron}, {Charton},
  {Chauvin}, {Claudi}, {Costille}, {De Caprio}, {de Boer}, {Delboulb{\'e}},
  {Desidera}, {Dominik}, {Downing}, {Dupuis}, {Fabron}, {Fantinel}, {Farisato},
  {Feautrier}, {Fedrigo}, {Fusco}, {Gigan}, {Ginski}, {Girard}, {Giro},
  {Gisler}, {Gluck}, {Gry}, {Henning}, {Hubin}, {Hugot}, {Incorvaia}, {Jaquet},
  {Kasper}, {Lagadec}, {Lagrange}, {Le Coroller}, {Le Mignant}, {Le Ruyet},
  {Lessio}, {Lizon}, {Llored}, {Lundin}, {Madec}, {Magnard}, {Marteaud},
  {Martinez}, {Maurel}, {M{\'e}nard}, {Mesa}, {M{\"o}ller-Nilsson}, {Moulin},
  {Moutou}, {Orign{\'e}}, {Parisot}, {Pavlov}, {Perret}, {Pragt}, {Puget},
  {Rabou}, {Ramos}, {Reess}, {Rigal}, {Rochat}, {Roelfsema}, {Rousset}, {Roux},
  {Saisse}, {Salasnich}, {Santambrogio}, {Scuderi}, {Segransan}, {Sevin},
  {Siebenmorgen}, {Soenke}, {Stadler}, {Suarez}, {Tiph{\`e}ne}, {Turatto},
  {Udry}, {Vakili}, {Waters}, {Weber}, {Wildi}, {Zins}, \&
  {Zurlo}}]{Beuzit2019}
{Beuzit}, J.~L., {Vigan}, A., {Mouillet}, D., {et~al.} 2019,
  \href{http://dx.doi.org/10.1051/0004-6361/201935251}{\color{blue}\aap},
  \href{https://ui.adsabs.harvard.edu/abs/2019A&A...631A.155B}{631, A155}

\bibitem[{Bradski(2000)}]{opencv_library}
Bradski, G. 2000, Dr. Dobb's Journal of Software Tools

\bibitem[{{Chakraborty} \& {Ge}(2004)}]{chakraborty2004}
{Chakraborty}, A. \& {Ge}, J. 2004,
  \href{http://dx.doi.org/10.1086/383287}{\color{blue}\aj},
  \href{https://ui.adsabs.harvard.edu/abs/2004AJ....127.2898C}{127, 2898}

\bibitem[{{Chauvin} {et~al.}(2017){Chauvin}, {Desidera}, {Lagrange}, {Vigan},
  {Gratton}, {Langlois}, {Bonnefoy}, {Beuzit}, {Feldt}, {Mouillet}, {Meyer},
  {Cheetham}, {Biller}, {Boccaletti}, {D'Orazi}, {Galicher}, {Hagelberg},
  {Maire}, {Mesa}, {Olofsson}, {Samland}, {Schmidt}, {Sissa}, {Bonavita},
  {Charnay}, {Cudel}, {Daemgen}, {Delorme}, {Janin-Potiron}, {Janson},
  {Keppler}, {Le Coroller}, {Ligi}, {Marleau}, {Messina}, {Molli{\`e}re},
  {Mordasini}, {M{\"u}ller}, {Peretti}, {Perrot}, {Rodet}, {Rouan}, {Zurlo},
  {Dominik}, {Henning}, {Menard}, {Schmid}, {Turatto}, {Udry}, {Vakili}, {Abe},
  {Antichi}, {Baruffolo}, {Baudoz}, {Baudrand}, {Blanchard}, {Bazzon}, {Buey},
  {Carbillet}, {Carle}, {Charton}, {Cascone}, {Claudi}, {Costille}, {Deboulbe},
  {De Caprio}, {Dohlen}, {Fantinel}, {Feautrier}, {Fusco}, {Gigan}, {Giro},
  {Gisler}, {Gluck}, {Hubin}, {Hugot}, {Jaquet}, {Kasper}, {Madec}, {Magnard},
  {Martinez}, {Maurel}, {Le Mignant}, {M{\"o}ller-Nilsson}, {Llored}, {Moulin},
  {Orign{\'e}}, {Pavlov}, {Perret}, {Petit}, {Pragt}, {Puget}, {Rabou},
  {Ramos}, {Rigal}, {Rochat}, {Roelfsema}, {Rousset}, {Roux}, {Salasnich},
  {Sauvage}, {Sevin}, {Soenke}, {Stadler}, {Suarez}, {Weber}, {Wildi},
  {Antoniucci}, {Augereau}, {Baudino}, {Brandner}, {Engler}, {Girard}, {Gry},
  {Kral}, {Kopytova}, {Lagadec}, {Milli}, {Moutou}, {Schlieder},
  {Szul{\'a}gyi}, {Thalmann}, \& {Wahhaj}}]{Chauvin2017}
{Chauvin}, G., {Desidera}, S., {Lagrange}, A.~M., {et~al.} 2017,
  \href{http://dx.doi.org/10.1051/0004-6361/201731152}{\color{blue}\aap},
  \href{https://ui.adsabs.harvard.edu/abs/2017A&A...605L...9C}{605, L9}

\bibitem[{{Currie} {et~al.}(2015){Currie}, {Cloutier}, {Brittain}, {Grady},
  {Burrows}, {Muto}, {Kenyon}, \& {Kuchner}}]{Currie2015}
{Currie}, T., {Cloutier}, R., {Brittain}, S., {et~al.} 2015,
  \href{http://dx.doi.org/10.1088/2041-8205/814/2/L27}{\color{blue}\apjl},
  \href{https://ui.adsabs.harvard.edu/abs/2015ApJ...814L..27C}{814, L27}

\bibitem[{{Currie} {et~al.}(2014){Currie}, {Muto}, {Kudo}, {Honda}, {Brandt},
  {Grady}, {Fukagawa}, {Burrows}, {Janson}, {Kuzuhara}, {McElwain}, {Follette},
  {Hashimoto}, {Henning}, {Kand ori}, {Kusakabe}, {Kwon}, {Mede}, {Morino},
  {Nishikawa}, {Pyo}, {Serabyn}, {Suenaga}, {Takahashi}, {Wisniewski}, \&
  {Tamura}}]{Currie2014}
{Currie}, T., {Muto}, T., {Kudo}, T., {et~al.} 2014,
  \href{http://dx.doi.org/10.1088/2041-8205/796/2/L30}{\color{blue}\apjl},
  \href{https://ui.adsabs.harvard.edu/abs/2014ApJ...796L..30C}{796, L30}

\bibitem[{{de Boer} {et~al.}(2016){de Boer}, {Salter}, {Benisty}, {Vigan},
  {Boccaletti}, {Pinilla}, {Ginski}, {Juhasz}, {Maire}, {Messina}, {Desidera},
  {Cheetham}, {Girard}, {Wahhaj}, {Langlois}, {Bonnefoy}, {Beuzit}, {Buenzli},
  {Chauvin}, {Dominik}, {Feldt}, {Gratton}, {Hagelberg}, {Isella}, {Janson},
  {Keller}, {Lagrange}, {Lannier}, {Menard}, {Mesa}, {Mouillet}, {Mugrauer},
  {Peretti}, {Perrot}, {Sissa}, {Snik}, {Vogt}, {Zurlo}, \& {SPHERE
  Consortium}}]{deBoer2016}
{de Boer}, J., {Salter}, G., {Benisty}, M., {et~al.} 2016,
  \href{http://dx.doi.org/10.1051/0004-6361/201629267}{\color{blue}\aap},
  \href{https://ui.adsabs.harvard.edu/abs/2016A&A...595A.114D}{595, A114}

\bibitem[{{de Leon} {et~al.}(2015){de Leon}, {Takami}, {Karr}, {Hashimoto},
  {Kudo}, {Sitko}, {Mayama}, {Kusakabe}, {Akiyama}, {Liu}, {Usuda}, {Abe},
  {Brand ner}, {Brandt}, {Carson}, {Currie}, {Egner}, {Feldt}, {Follette},
  {Grady}, {Goto}, {Guyon}, {Hayano}, {Hayashi}, {Hayashi}, {Henning},
  {Hodapp}, {Ishii}, {Iye}, {Janson}, {Kand ori}, {Knapp}, {Kuzuhara}, {Kwon},
  {Matsuo}, {McElwain}, {Miyama}, {Morino}, {Moro-Martin}, {Nishimura}, {Pyo},
  {Serabyn}, {Suenaga}, {Suto}, {Suzuki}, {Takahashi}, {Takato}, {Terada},
  {Thalmann}, {Tomono}, {Turner}, {Watanabe}, {Wisniewski}, {Yamada}, {Takami},
  \& {Tamura}}]{deLeon2015}
{de Leon}, J., {Takami}, M., {Karr}, J.~L., {et~al.} 2015,
  \href{http://dx.doi.org/10.1088/2041-8205/806/1/L10}{\color{blue}\apjl},
  \href{https://ui.adsabs.harvard.edu/abs/2015ApJ...806L..10D}{806, L10}

\bibitem[{Dryden \& Mardia(2016)}]{Dryden2016}
Dryden, I.~L. \& Mardia, K.~V. 2016, Statistical shape analysis: with
  applications in R, Vol. 995 (John Wiley \& Sons)

\bibitem[{{Follette} {et~al.}(2017){Follette}, {Rameau}, {Dong}, {Pueyo},
  {Close}, {Duch{\^e}ne}, {Fung}, {Leonard}, {Macintosh}, {Males}, {Marois},
  {Millar-Blanchaer}, {Morzinski}, {Mullen}, {Perrin}, {Spiro}, {Wang},
  {Ammons}, {Bailey}, {Barman}, {Bulger}, {Chilcote}, {Cotten}, {De Rosa},
  {Doyon}, {Fitzgerald}, {Goodsell}, {Graham}, {Greenbaum}, {Hibon}, {Hung},
  {Ingraham}, {Kalas}, {Konopacky}, {Larkin}, {Maire}, {Marchis}, {Metchev},
  {Nielsen}, {Oppenheimer}, {Palmer}, {Patience}, {Poyneer}, {Rajan},
  {Rantakyr{\"o}}, {Savransky}, {Schneider}, {Sivaramakrishnan}, {Song},
  {Soummer}, {Thomas}, {Vega}, {Wallace}, {Ward-Duong}, {Wiktorowicz}, \&
  {Wolff}}]{Follette2017}
{Follette}, K.~B., {Rameau}, J., {Dong}, R., {et~al.} 2017,
  \href{http://dx.doi.org/10.3847/1538-3881/aa6d85}{\color{blue}\aj},
  \href{https://ui.adsabs.harvard.edu/abs/2017AJ....153..264F}{153, 264}

\bibitem[{{Garufi} {et~al.}(2016){Garufi}, {Quanz}, {Schmid}, {Mulders},
  {Avenhaus}, {Boccaletti}, {Ginski}, {Langlois}, {Stolker}, {Augereau},
  {Benisty}, {Lopez}, {Dominik}, {Gratton}, {Henning}, {Janson}, {M{\'e}nard},
  {Meyer}, {Pinte}, {Sissa}, {Vigan}, {Zurlo}, {Bazzon}, {Buenzli}, {Bonnefoy},
  {Brandner}, {Chauvin}, {Cheetham}, {Cudel}, {Desidera}, {Feldt}, {Galicher},
  {Kasper}, {Lagrange}, {Lannier}, {Maire}, {Mesa}, {Mouillet}, {Peretti},
  {Perrot}, {Salter}, \& {Wildi}}]{Garufi2016}
{Garufi}, A., {Quanz}, S.~P., {Schmid}, H.~M., {et~al.} 2016,
  \href{http://dx.doi.org/10.1051/0004-6361/201527940}{\color{blue}\aap},
  \href{https://ui.adsabs.harvard.edu/abs/2016A&A...588A...8G}{588, A8}

\bibitem[{{Ginski} {et~al.}(2021){Ginski}, {Facchini}, {Huang}, {Benisty},
  {Vaendel}, {Stapper}, {Dominik}, {Bae}, {M{\'e}nard}, {Muro-Arena},
  {Hogerheijde}, {McClure}, {van Holstein}, {Birnstiel}, {Boehler}, {Bohn},
  {Flock}, {Mamajek}, {Manara}, {Pinilla}, {Pinte}, \& {Ribas}}]{Ginski2021}
{Ginski}, C., {Facchini}, S., {Huang}, J., {et~al.} 2021,
  \href{http://dx.doi.org/10.3847/2041-8213/abdf57}{\color{blue}\apjl},
  \href{https://ui.adsabs.harvard.edu/abs/2021ApJ...908L..25G}{908, L25}

\bibitem[{{Ginski} {et~al.}(2016){Ginski}, {Stolker}, {Pinilla}, {Dominik},
  {Boccaletti}, {de Boer}, {Benisty}, {Biller}, {Feldt}, {Garufi}, {Keller},
  {Kenworthy}, {Maire}, {M{\'e}nard}, {Mesa}, {Milli}, {Min}, {Pinte}, {Quanz},
  {van Boekel}, {Bonnefoy}, {Chauvin}, {Desidera}, {Gratton}, {Girard},
  {Keppler}, {Kopytova}, {Lagrange}, {Langlois}, {Rouan}, \&
  {Vigan}}]{Ginski2016}
{Ginski}, C., {Stolker}, T., {Pinilla}, P., {et~al.} 2016,
  \href{http://dx.doi.org/10.1051/0004-6361/201629265}{\color{blue}\aap},
  \href{https://ui.adsabs.harvard.edu/abs/2016A&A...595A.112G}{595, A112}

\bibitem[{Harris {et~al.}(2020)Harris, Millman, van~der Walt, Gommers,
  Virtanen, Cournapeau, Wieser, Taylor, Berg, Smith, Kern, Picus, Hoyer, van
  Kerkwijk, Brett, Haldane, del R{\'{i}}o, Wiebe, Peterson,
  G{\'{e}}rard-Marchant, Sheppard, Reddy, Weckesser, Abbasi, Gohlke, \&
  Oliphant}]{Harris2020}
Harris, C.~R., Millman, K.~J., van~der Walt, S.~J., {et~al.} 2020,
  \href{http://dx.doi.org/10.1038/s41586-020-2649-2}{\color{blue}Nature}, 585,
  585

\bibitem[{Hunter(2007)}]{Hunter2007}
Hunter, J.~D. 2007,
  \href{http://dx.doi.org/10.1109/MCSE.2007.55}{\color{blue}Computing in
  Science \& Engineering}, 9, 9

\bibitem[{{Jeffers} {et~al.}(2014){Jeffers}, {Min}, {Canovas}, {Rodenhuis}, \&
  {Keller}}]{Jeffers2014}
{Jeffers}, S.~V., {Min}, M., {Canovas}, H., {Rodenhuis}, M., \& {Keller}, C.~U.
  2014, \href{http://dx.doi.org/10.1051/0004-6361/201220186}{\color{blue}\aap},
  \href{https://ui.adsabs.harvard.edu/abs/2014A&A...561A..23J}{561, A23}

\bibitem[{{Kuhn} {et~al.}(2001){Kuhn}, {Potter}, \& {Parise}}]{Kuhn2001}
{Kuhn}, J.~R., {Potter}, D., \& {Parise}, B. 2001,
  \href{http://dx.doi.org/10.1086/320686}{\color{blue}\apjl},
  \href{https://ui.adsabs.harvard.edu/abs/2001ApJ...553L.189K}{553, L189}

\bibitem[{{Lafreni{\`e}re} {et~al.}(2007){Lafreni{\`e}re}, {Marois}, {Doyon},
  {Nadeau}, \& {Artigau}}]{Lafreniere2007}
{Lafreni{\`e}re}, D., {Marois}, C., {Doyon}, R., {Nadeau}, D., \& {Artigau},
  {\'E}. 2007, \href{http://dx.doi.org/10.1086/513180}{\color{blue}\apj},
  \href{https://ui.adsabs.harvard.edu/abs/2007ApJ...660..770L}{660, 770}

\bibitem[{{Laws} {et~al.}(2020){Laws}, {Harries}, {Setterholm}, {Monnier},
  {Rich}, {Aarnio}, {Adams}, {Andrews}, {Bae}, {Calvet}, {Espaillat},
  {Hartmann}, {Hinkley}, {Isella}, {Kraus}, {Wilner}, \& {Zhu}}]{Laws2020}
{Laws}, A. S.~E., {Harries}, T.~J., {Setterholm}, B.~R., {et~al.} 2020,
  \href{http://dx.doi.org/10.3847/1538-4357/ab59e2}{\color{blue}\apj},
  \href{https://ui.adsabs.harvard.edu/abs/2020ApJ...888....7L}{888, 7}

\bibitem[{{Macintosh} {et~al.}(2015){Macintosh}, {Graham}, {Barman}, {De Rosa},
  {Konopacky}, {Marley}, {Marois}, {Nielsen}, {Pueyo}, {Rajan}, {Rameau},
  {Saumon}, {Wang}, {Patience}, {Ammons}, {Arriaga}, {Artigau}, {Beckwith},
  {Brewster}, {Bruzzone}, {Bulger}, {Burningham}, {Burrows}, {Chen}, {Chiang},
  {Chilcote}, {Dawson}, {Dong}, {Doyon}, {Draper}, {Duch{\^e}ne}, {Esposito},
  {Fabrycky}, {Fitzgerald}, {Follette}, {Fortney}, {Gerard}, {Goodsell},
  {Greenbaum}, {Hibon}, {Hinkley}, {Cotten}, {Hung}, {Ingraham},
  {Johnson-Groh}, {Kalas}, {Lafreniere}, {Larkin}, {Lee}, {Line}, {Long},
  {Maire}, {Marchis}, {Matthews}, {Max}, {Metchev}, {Millar-Blanchaer},
  {Mittal}, {Morley}, {Morzinski}, {Murray-Clay}, {Oppenheimer}, {Palmer},
  {Patel}, {Perrin}, {Poyneer}, {Rafikov}, {Rantakyr{\"o}}, {Rice}, {Rojo},
  {Rudy}, {Ruffio}, {Ruiz}, {Sadakuni}, {Saddlemyer}, {Salama}, {Savransky},
  {Schneider}, {Sivaramakrishnan}, {Song}, {Soummer}, {Thomas}, {Vasisht},
  {Wallace}, {Ward-Duong}, {Wiktorowicz}, {Wolff}, \&
  {Zuckerman}}]{Macintosh2015}
{Macintosh}, B., {Graham}, J.~R., {Barman}, T., {et~al.} 2015,
  \href{http://dx.doi.org/10.1126/science.aac5891}{\color{blue}Science},
  \href{https://ui.adsabs.harvard.edu/abs/2015Sci...350...64M}{350, 64}

\bibitem[{{Macintosh} {et~al.}(2014){Macintosh}, {Graham}, {Ingraham},
  {Konopacky}, {Marois}, {Perrin}, {Poyneer}, {Bauman}, {Barman}, {Burrows},
  {Cardwell}, {Chilcote}, {De Rosa}, {Dillon}, {Doyon}, {Dunn}, {Erikson},
  {Fitzgerald}, {Gavel}, {Goodsell}, {Hartung}, {Hibon}, {Kalas}, {Larkin},
  {Maire}, {Marchis}, {Marley}, {McBride}, {Millar-Blanchaer}, {Morzinski},
  {Norton}, {Oppenheimer}, {Palmer}, {Patience}, {Pueyo}, {Rantakyro},
  {Sadakuni}, {Saddlemyer}, {Savransky}, {Serio}, {Soummer},
  {Sivaramakrishnan}, {Song}, {Thomas}, {Wallace}, {Wiktorowicz}, \&
  {Wolff}}]{Macintosh2014}
{Macintosh}, B., {Graham}, J.~R., {Ingraham}, P., {et~al.} 2014,
  \href{http://dx.doi.org/10.1073/pnas.1304215111}{\color{blue}Proceedings of
  the National Academy of Science},
  \href{https://ui.adsabs.harvard.edu/abs/2014PNAS..11112661M}{111, 12661}

\bibitem[{{Marois} {et~al.}(2006){Marois}, {Lafreni{\`e}re}, {Doyon},
  {Macintosh}, \& {Nadeau}}]{Marois2006}
{Marois}, C., {Lafreni{\`e}re}, D., {Doyon}, R., {Macintosh}, B., \& {Nadeau},
  D. 2006, \href{http://dx.doi.org/10.1086/500401}{\color{blue}\apj},
  \href{https://ui.adsabs.harvard.edu/abs/2006ApJ...641..556M}{641, 556}

\bibitem[{{Marois} {et~al.}(2008){Marois}, {Macintosh}, {Barman}, {Zuckerman},
  {Song}, {Patience}, {Lafreni{\`e}re}, \& {Doyon}}]{Marois2008}
{Marois}, C., {Macintosh}, B., {Barman}, T., {et~al.} 2008,
  \href{http://dx.doi.org/10.1126/science.1166585}{\color{blue}Science},
  \href{https://ui.adsabs.harvard.edu/abs/2008Sci...322.1348M}{322, 1348}

\bibitem[{{Marois} {et~al.}(2010){Marois}, {Zuckerman}, {Konopacky},
  {Macintosh}, \& {Barman}}]{Marois2010}
{Marois}, C., {Zuckerman}, B., {Konopacky}, Q.~M., {Macintosh}, B., \&
  {Barman}, T. 2010,
  \href{http://dx.doi.org/10.1038/nature09684}{\color{blue}\nat},
  \href{https://ui.adsabs.harvard.edu/abs/2010Natur.468.1080M}{468, 1080}

\bibitem[{{Milli} {et~al.}(2012){Milli}, {Mouillet}, {Lagrange}, {Boccaletti},
  {Mawet}, {Chauvin}, \& {Bonnefoy}}]{Milli2012}
{Milli}, J., {Mouillet}, D., {Lagrange}, A.~M., {et~al.} 2012,
  \href{http://dx.doi.org/10.1051/0004-6361/201219687}{\color{blue}\aap},
  \href{https://ui.adsabs.harvard.edu/abs/2012A&A...545A.111M}{545, A111}

\bibitem[{{Min} {et~al.}(2005){Min}, {Hovenier}, \& {de Koter}}]{Min2005}
{Min}, M., {Hovenier}, J.~W., \& {de Koter}, A. 2005,
  \href{http://dx.doi.org/10.1051/0004-6361:20041920}{\color{blue}\aap},
  \href{https://ui.adsabs.harvard.edu/abs/2005A&A...432..909M}{432, 909}

\bibitem[{{Monnier} {et~al.}(2019){Monnier}, {Harries}, {Bae}, {Setterholm},
  {Laws}, {Aarnio}, {Adams}, {Andrews}, {Calvet}, {Espaillat}, {Hartmann},
  {Kraus}, {McClure}, {Miller}, {Oppenheimer}, {Wilner}, \&
  {Zhu}}]{Monnier2019}
{Monnier}, J.~D., {Harries}, T.~J., {Bae}, J., {et~al.} 2019,
  \href{http://dx.doi.org/10.3847/1538-4357/aafe87}{\color{blue}\apj},
  \href{https://ui.adsabs.harvard.edu/abs/2019ApJ...872..122M}{872, 122}

\bibitem[{{Murakawa}(2010)}]{Murakawa2010}
{Murakawa}, K. 2010,
  \href{http://dx.doi.org/10.1051/0004-6361/201014159}{\color{blue}\aap},
  \href{https://ui.adsabs.harvard.edu/abs/2010A&A...518A..63M}{518, A63}

\bibitem[{{Muro-Arena} {et~al.}(2020){Muro-Arena}, {Benisty}, {Ginski},
  {Dominik}, {Facchini}, {Villenave}, {van Boekel}, {Chauvin}, {Garufi},
  {Henning}, {Janson}, {Keppler}, {Matter}, {M{\'e}nard}, {Stolker}, {Zurlo},
  {Blanchard}, {Maurel}, {Moeller-Nilsson}, {Petit}, {Roux}, {Sevin}, \&
  {Wildi}}]{MuroArena2020}
{Muro-Arena}, G.~A., {Benisty}, M., {Ginski}, C., {et~al.} 2020,
  \href{http://dx.doi.org/10.1051/0004-6361/201936509}{\color{blue}\aap},
  \href{https://ui.adsabs.harvard.edu/abs/2020A&A...635A.121M}{635, A121}

\bibitem[{{Pairet} {et~al.}(2018){Pairet}, {Cantalloube}, \&
  {Jacques}}]{Pairet2018}
{Pairet}, B., {Cantalloube}, F., \& {Jacques}, L. 2018,
  \href{https://ui.adsabs.harvard.edu/abs/2018arXiv181201333P}{arXiv e-prints,
  arXiv:1812.01333}

\bibitem[{{Pairet} {et~al.}(2021){Pairet}, {Cantalloube}, \&
  {Jacques}}]{Pairet2021}
{Pairet}, B., {Cantalloube}, F., \& {Jacques}, L. 2021,
  \href{http://dx.doi.org/10.1093/mnras/stab607}{\color{blue}\mnras},
  \href{https://ui.adsabs.harvard.edu/abs/2021MNRAS.503.3724P}{503, 3724}

\bibitem[{{Perrot} {et~al.}(2016){Perrot}, {Boccaletti}, {Pantin}, {Augereau},
  {Lagrange}, {Galicher}, {Maire}, {Mazoyer}, {Milli}, {Rousset}, {Gratton},
  {Bonnefoy}, {Brandner}, {Buenzli}, {Langlois}, {Lannier}, {Mesa}, {Peretti},
  {Salter}, {Sissa}, {Chauvin}, {Desidera}, {Feldt}, {Vigan}, {Di Folco},
  {Dutrey}, {P{\'e}ricaud}, {Baudoz}, {Benisty}, {De Boer}, {Garufi}, {Girard},
  {Menard}, {Olofsson}, {Quanz}, {Mouillet}, {Christiaens}, {Casassus},
  {Beuzit}, {Blanchard}, {Carle}, {Fusco}, {Giro}, {Hubin}, {Maurel},
  {Moeller-Nilsson}, {Sevin}, \& {Weber}}]{Perrot2016}
{Perrot}, C., {Boccaletti}, A., {Pantin}, E., {et~al.} 2016,
  \href{http://dx.doi.org/10.1051/0004-6361/201628396}{\color{blue}\aap},
  \href{https://ui.adsabs.harvard.edu/abs/2016A&A...590L...7P}{590, L7}

\bibitem[{{Pineda} {et~al.}(2019){Pineda}, {Szul{\'a}gyi}, {Quanz}, {van
  Dishoeck}, {Garufi}, {Meru}, {Mulders}, {Testi}, {Meyer}, \&
  {Reggiani}}]{pineda2019}
{Pineda}, J.~E., {Szul{\'a}gyi}, J., {Quanz}, S.~P., {et~al.} 2019,
  \href{http://dx.doi.org/10.3847/1538-4357/aaf389}{\color{blue}\apj},
  \href{https://ui.adsabs.harvard.edu/abs/2019ApJ...871...48P}{871, 48}

\bibitem[{{Pohl} {et~al.}(2017){Pohl}, {Benisty}, {Pinilla}, {Ginski}, {de
  Boer}, {Avenhaus}, {Henning}, {Zurlo}, {Boccaletti}, {Augereau}, {Birnstiel},
  {Dominik}, {Facchini}, {Fedele}, {Janson}, {Keppler}, {Kral}, {Langlois},
  {Ligi}, {Maire}, {M{\'e}nard}, {Meyer}, {Pinte}, {Quanz}, {Sauvage},
  {Sezestre}, {Stolker}, {Szul{\'a}gyi}, {van Boekel}, {van der Plas},
  {Villenave}, {Baruffolo}, {Baudoz}, {Le Mignant}, {Maurel}, {Ramos}, \&
  {Weber}}]{Pohl2017}
{Pohl}, A., {Benisty}, M., {Pinilla}, P., {et~al.} 2017,
  \href{http://dx.doi.org/10.3847/1538-4357/aa94c2}{\color{blue}\apj},
  \href{https://ui.adsabs.harvard.edu/abs/2017ApJ...850...52P}{850, 52}

\bibitem[{{Quanz} {et~al.}(2015){Quanz}, {Amara}, {Meyer}, {Girard},
  {Kenworthy}, \& {Kasper}}]{quanz2015}
{Quanz}, S.~P., {Amara}, A., {Meyer}, M.~R., {et~al.} 2015,
  \href{http://dx.doi.org/10.1088/0004-637X/807/1/64}{\color{blue}\apj},
  \href{https://ui.adsabs.harvard.edu/abs/2015ApJ...807...64Q}{807, 64}

\bibitem[{{Quanz} {et~al.}(2013){Quanz}, {Amara}, {Meyer}, {Kenworthy},
  {Kasper}, \& {Girard}}]{quanz2013}
{Quanz}, S.~P., {Amara}, A., {Meyer}, M.~R., {et~al.} 2013,
  \href{http://dx.doi.org/10.1088/2041-8205/766/1/L1}{\color{blue}\apjl},
  \href{https://ui.adsabs.harvard.edu/abs/2013ApJ...766L...1Q}{766, L1}

\bibitem[{{Rameau} {et~al.}(2017){Rameau}, {Follette}, {Pueyo}, {Marois},
  {Macintosh}, {Millar-Blanchaer}, {Wang}, {Vega}, {Doyon}, {Lafreni{\`e}re},
  {Nielsen}, {Bailey}, {Chilcote}, {Close}, {Esposito}, {Males}, {Metchev},
  {Morzinski}, {Ruffio}, {Wolff}, {Ammons}, {Barman}, {Bulger}, {Cotten}, {De
  Rosa}, {Duchene}, {Fitzgerald}, {Goodsell}, {Graham}, {Greenbaum}, {Hibon},
  {Hung}, {Ingraham}, {Kalas}, {Konopacky}, {Larkin}, {Maire}, {Marchis},
  {Oppenheimer}, {Palmer}, {Patience}, {Perrin}, {Poyneer}, {Rajan},
  {Rantakyr{\"o}}, {Marley}, {Savransky}, {Schneider}, {Sivaramakrishnan},
  {Song}, {Soummer}, {Thomas}, {Wallace}, {Ward-Duong}, \&
  {Wiktorowicz}}]{Rameau2017}
{Rameau}, J., {Follette}, K.~B., {Pueyo}, L., {et~al.} 2017,
  \href{http://dx.doi.org/10.3847/1538-3881/aa6cae}{\color{blue}\aj},
  \href{https://ui.adsabs.harvard.edu/abs/2017AJ....153..244R}{153, 244}

\bibitem[{{Ren} {et~al.}(2018){Ren}, {Pueyo}, {Zhu}, {Debes}, \&
  {Duch{\^e}ne}}]{Ren2018}
{Ren}, B., {Pueyo}, L., {Zhu}, G.~B., {Debes}, J., \& {Duch{\^e}ne}, G. 2018,
  \href{http://dx.doi.org/10.3847/1538-4357/aaa1f2}{\color{blue}\apj},
  \href{https://ui.adsabs.harvard.edu/abs/2018ApJ...852..104R}{852, 104}

\bibitem[{{Sissa} {et~al.}(2018){Sissa}, {Gratton}, {Garufi}, {Rigliaco},
  {Zurlo}, {Mesa}, {Langlois}, {de Boer}, {Desidera}, {Ginski}, {Lagrange},
  {Maire}, {Vigan}, {Dima}, {Antichi}, {Baruffolo}, {Bazzon}, {Benisty},
  {Beuzit}, {Biller}, {Boccaletti}, {Bonavita}, {Bonnefoy}, {Brandner},
  {Bruno}, {Buenzli}, {Cascone}, {Chauvin}, {Cheetham}, {Claudi}, {Cudel}, {De
  Caprio}, {Dominik}, {Fantinel}, {Farisato}, {Feldt}, {Fontanive}, {Galicher},
  {Giro}, {Hagelberg}, {Incorvaia}, {Janson}, {Kasper}, {Keppler}, {Kopytova},
  {Lagadec}, {Lannier}, {Lazzoni}, {LeCoroller}, {Lessio}, {Ligi}, {Marzari},
  {Menard}, {Meyer}, {Mouillet}, {Peretti}, {Perrot}, {Potiron}, {Rouan},
  {Salasnich}, {Salter}, {Samland}, {Schmidt}, {Scuderi}, \&
  {Wildi}}]{Sissa2018}
{Sissa}, E., {Gratton}, R., {Garufi}, A., {et~al.} 2018,
  \href{http://dx.doi.org/10.1051/0004-6361/201732332}{\color{blue}\aap},
  \href{https://ui.adsabs.harvard.edu/abs/2018A&A...619A.160S}{619, A160}

\bibitem[{{Stolker} {et~al.}(2019){Stolker}, {Bonse}, {Quanz}, {Amara},
  {Cugno}, {Bohn}, \& {Boehle}}]{Stolker2019}
{Stolker}, T., {Bonse}, M.~J., {Quanz}, S.~P., {et~al.} 2019,
  \href{http://dx.doi.org/10.1051/0004-6361/201834136}{\color{blue}\aap},
  \href{https://ui.adsabs.harvard.edu/abs/2019A&A...621A..59S}{621, A59}

\bibitem[{{Stolker} {et~al.}(2016){Stolker}, {Dominik}, {Avenhaus}, {Min}, {de
  Boer}, {Ginski}, {Schmid}, {Juhasz}, {Bazzon}, {Waters}, {Garufi},
  {Augereau}, {Benisty}, {Boccaletti}, {Henning}, {Langlois}, {Maire},
  {M{\'e}nard}, {Meyer}, {Pinte}, {Quanz}, {Thalmann}, {Beuzit}, {Carbillet},
  {Costille}, {Dohlen}, {Feldt}, {Gisler}, {Mouillet}, {Pavlov}, {Perret},
  {Petit}, {Pragt}, {Rochat}, {Roelfsema}, {Salasnich}, {Soenke}, \&
  {Wildi}}]{Stolker2016}
{Stolker}, T., {Dominik}, C., {Avenhaus}, H., {et~al.} 2016,
  \href{http://dx.doi.org/10.1051/0004-6361/201528039}{\color{blue}\aap},
  \href{https://ui.adsabs.harvard.edu/abs/2016A&A...595A.113S}{595, A113}

\bibitem[{{Stolker} {et~al.}(2017){Stolker}, {Min}, {Stam}, {Molli{\`e}re},
  {Dominik}, \& {Waters}}]{Stolker2017}
{Stolker}, T., {Min}, M., {Stam}, D.~M., {et~al.} 2017,
  \href{http://dx.doi.org/10.1051/0004-6361/201730780}{\color{blue}\aap},
  \href{https://ui.adsabs.harvard.edu/abs/2017A&A...607A..42S}{607, A42}

\bibitem[{{Tazaki} {et~al.}(2019){Tazaki}, {Tanaka}, {Muto}, {Kataoka}, \&
  {Okuzumi}}]{Tazaki2019}
{Tazaki}, R., {Tanaka}, H., {Muto}, T., {Kataoka}, A., \& {Okuzumi}, S. 2019,
  \href{http://dx.doi.org/10.1093/mnras/stz662}{\color{blue}\mnras},
  \href{https://ui.adsabs.harvard.edu/abs/2019MNRAS.485.4951T}{485, 4951}

\bibitem[{{Uyama} {et~al.}(2020){Uyama}, {Currie}, {Christiaens}, {Bae},
  {Muto}, {Takahashi}, {Tazaki}, {Ygouf}, {Kasdin}, {Groff}, {Brandt},
  {Chilcote}, {Hayashi}, {McElwain}, {Guyon}, {Lozi}, {Jovanovic},
  {Martinache}, {Kudo}, {Tamura}, {Akiyama}, {Beichman}, {Grady}, {Knapp},
  {Kwon}, {Sitko}, {Takami}, {Wagner}, {Wisniewski}, \& {Yang}}]{Uyama2020}
{Uyama}, T., {Currie}, T., {Christiaens}, V., {et~al.} 2020,
  \href{http://dx.doi.org/10.3847/1538-4357/aba8f6}{\color{blue}\apj},
  \href{https://ui.adsabs.harvard.edu/abs/2020ApJ...900..135U}{900, 135}

\bibitem[{{van Boekel} {et~al.}(2017){van Boekel}, {Henning}, {Menu}, {de
  Boer}, {Langlois}, {M{\"u}ller}, {Avenhaus}, {Boccaletti}, {Schmid},
  {Thalmann}, {Benisty}, {Dominik}, {Ginski}, {Girard}, {Gisler}, {Lobo Gomes},
  {Menard}, {Min}, {Pavlov}, {Pohl}, {Quanz}, {Rabou}, {Roelfsema}, {Sauvage},
  {Teague}, {Wildi}, \& {Zurlo}}]{vanBoekel2017}
{van Boekel}, R., {Henning}, T., {Menu}, J., {et~al.} 2017,
  \href{http://dx.doi.org/10.3847/1538-4357/aa5d68}{\color{blue}\apj},
  \href{https://ui.adsabs.harvard.edu/abs/2017ApJ...837..132V}{837, 132}

\bibitem[{{van Holstein} {et~al.}(2020){van Holstein}, {Girard}, {de Boer},
  {Snik}, {Milli}, {Stam}, {Ginski}, {Mouillet}, {Wahhaj}, {Schmid}, {Keller},
  {Langlois}, {Dohlen}, {Vigan}, {Pohl}, {Carbillet}, {Fantinel}, {Maurel},
  {Orign{\'e}}, {Petit}, {Ramos}, {Rigal}, {Sevin}, {Boccaletti}, {Le
  Coroller}, {Dominik}, {Henning}, {Lagadec}, {M{\'e}nard}, {Turatto}, {Udry},
  {Chauvin}, {Feldt}, \& {Beuzit}}]{vanHolstein2020}
{van Holstein}, R.~G., {Girard}, J.~H., {de Boer}, J., {et~al.} 2020,
  \href{http://dx.doi.org/10.1051/0004-6361/201834996}{\color{blue}\aap},
  \href{https://ui.adsabs.harvard.edu/abs/2020A&A...633A..64V}{633, A64}

\bibitem[{{van Holstein} {et~al.}(2017){van Holstein}, {Snik}, {Girard}, {de
  Boer}, {Ginski}, {Keller}, {Stam}, {Beuzit}, {Mouillet}, {Kasper},
  {Langlois}, {Zurlo}, {de Kok}, \& {Vigan}}]{vanHolstein2017}
{van Holstein}, R.~G., {Snik}, F., {Girard}, J.~H., {et~al.} 2017, in Society
  of Photo-Optical Instrumentation Engineers (SPIE) Conference Series, Vol.
  10400, Society of Photo-Optical Instrumentation Engineers (SPIE) Conference
  Series, \href{https://ui.adsabs.harvard.edu/abs/2017SPIE10400E..15V}{1040015}

\bibitem[{{van Holstein} {et~al.}(2021){van Holstein}, {Stolker},
  {Jensen-Clem}, {Ginski}, {Milli}, {de Boer}, {Girard}, {Wahhaj}, {Bohn},
  {Millar-Blanchaer}, {Benisty}, {Bonnefoy}, {Chauvin}, {Dominik}, {Hinkley},
  {Keller}, {Keppler}, {Langlois}, {Marino}, {M{\'e}nard}, {Perrot}, {Schmidt},
  {Vigan}, {Zurlo}, \& {Snik}}]{vanHolstein2021}
{van Holstein}, R.~G., {Stolker}, T., {Jensen-Clem}, R., {et~al.} 2021,
  \href{http://dx.doi.org/10.1051/0004-6361/202039290}{\color{blue}\aap},
  \href{https://ui.adsabs.harvard.edu/abs/2021A&A...647A..21V}{647, A21}

\bibitem[{Virtanen {et~al.}(2020)Virtanen, Gommers, Oliphant, Haberland, Reddy,
  Cournapeau, Burovski, Peterson, Weckesser, Bright, {van der Walt}, Brett,
  Wilson, Millman, Mayorov, Nelson, Jones, Kern, Larson, Carey, Polat, Feng,
  Moore, {VanderPlas}, Laxalde, Perktold, Cimrman, Henriksen, Quintero, Harris,
  Archibald, Ribeiro, Pedregosa, {van Mulbregt}, \& {SciPy 1.0
  Contributors}}]{2020SciPy-NMeth}
Virtanen, P., Gommers, R., Oliphant, T.~E., {et~al.} 2020,
  \href{http://dx.doi.org/10.1038/s41592-019-0686-2}{\color{blue}Nature
  Methods}, \href{https://rdcu.be/b08Wh}{17, 261}

\bibitem[{{Wang} {et~al.}(2015){Wang}, {Ruffio}, {De Rosa}, {Aguilar}, {Wolff},
  \& {Pueyo}}]{Wang2015}
{Wang}, J.~J., {Ruffio}, J.-B., {De Rosa}, R.~J., {et~al.} 2015, {pyKLIP: PSF
  Subtraction for Exoplanets and Disks}, Astrophysics Source Code Library,
  record ascl:1506.001

\end{thebibliography}

\begin{appendix}
\section{Field rotation for different declinations and latitudes}
\label{app:field_rotation}
\begin{figure}[h]
    \centering
    \includegraphics[width=0.5\textwidth]{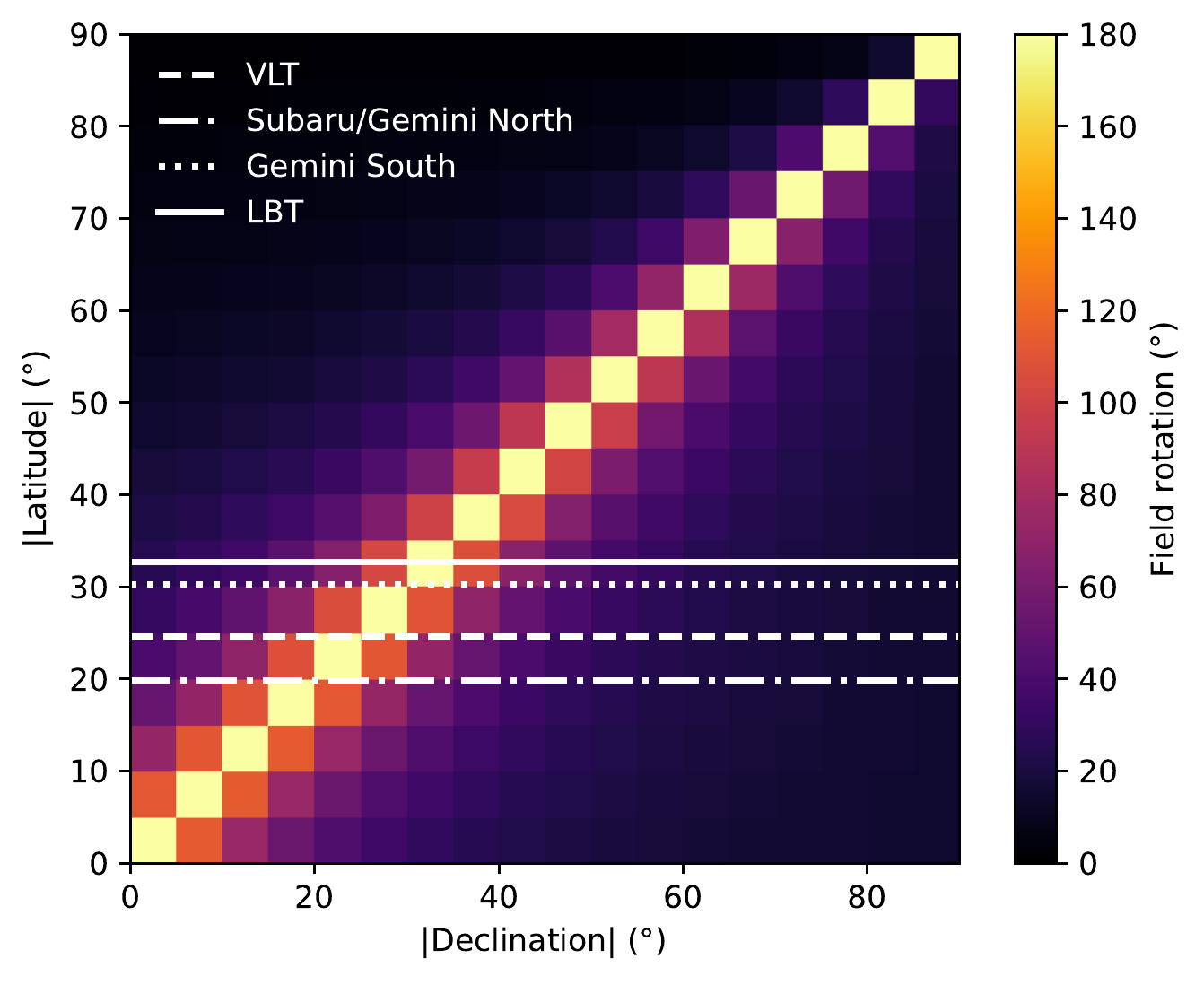}
    \caption{The absolute latitude of the telescope plotted against the absolute declination of the object. The colors represent the total amount of field rotation for a one hour observation. The lines are the latitudes of the major direct imaging facilities.}
    \label{fig:field_rotation}
\end{figure}

Figure \ref{fig:field_rotation} presents the amount of field rotation for different latitudes of the telescope and declinations of the target. When the declination and latitude are the same, the    is 90$\degree$ and the object has a total of 180$\degree$ of field rotation, which reduces for lower elevations. In most cases the amount of field rotation for different elevations is the same, especially for the major direct imaging facilities. For high latitudes the amount of field rotation reduces. Hence, in the case of high latitudes, the results shown in this work are optimistic because of much less field rotation. However, at these high latitudes ADI in general would be inefficient.

\end{appendix}


\end{document}